\documentclass[prd,aps,preprintnumbers,nofootinbib,twocolumn]{revtex4}

\usepackage{amscd}
\usepackage{amsmath, bbm}

\usepackage{amsfonts}
\usepackage{amssymb}
\usepackage{slashed}
\usepackage{multirow}
\usepackage{array}
\usepackage{xcolor}
\usepackage{hyperref}

\newcommand{\Q}{\mathbb{Q}}
\newcommand{\R}{\mathbb{R}}
\newcommand{\C}{\mathbb{C}}
\newcommand{\Z}{\mathbb{Z}}
\begin{document} 
\title{Solving local anomaly equations in gauge-rank extensions of the Standard Model}
\author{B C Allanach} \email{B.C.Allanach@damtp.cam.ac.uk} \affiliation{DAMTP, University of Cambridge, Wilberforce Road, Cambridge, 
CB3 0WA, United Kingdom}
\author{Ben Gripaios} \email{gripaios@hep.phy.cam.ac.uk} \affiliation{Cavendish Laboratory, University of Cambridge, J.J. Thomson
Avenue, Cambridge, CB3 0HE, United Kingdom}
\author{Joseph Tooby-Smith} \email{jss85@cam.ac.uk} \affiliation{Cavendish Laboratory, University of Cambridge, J.J. Thomson
Avenue, Cambridge, CB3 0HE, United Kingdom}
\begin{abstract}
We consider local (or perturbative) gauge anomalies in models which
  extend the rank of the Standard Model (SM) gauge group and the
  chiral fermion content only by $n$ SM singlets.
  We give a general solution to the 
  anomaly cancellation
  conditions (ACCs)
  of an additional $U(1)$ subgroup
  for the ACCs that involve only SM fermions and we examine whether 
  a corresponding solution exists for the remaining ACCs.
  We show
  that a solution to
the remaining ACCs always exists for $n \geq 5$ in the family
non-universal case or $n \geq 3$ in the family-universal case.
In the special case where only a single family carries non-vanishing charges, 
we find a general solution to all ACCs, for any value of $n$.
\end{abstract}

\maketitle
\section{Introduction}\label{sec:Intro}
Countless gauge extensions of the SM have been constructed in the
  literature. Most of these involve an increase in rank with respect
  to the usual SM
  gauge group,\footnote{An obvious, rare, exception is
    the $SU(5)$ GUT.} so there is usually at least one
  additional, non-linearly realised $U(1)$ subgroup in the extension.
But if one wishes to tinker with the gauge structure of
the SM in this way, one should check that basic requirements of
locality and unitarity of the theory are not violated via anomalies
in the fermionic path-integral. 

Our goal here is to carry out the necessary due diligence\footnote{It is well known that local anomalies place non-trivial constraints upon representations of chiral
  fermions. For example, in the SM, if we allow the hypercharges of the
  fermions to vary over the reals, the combination of gauge and
  gravitational anomaly cancellation implies that the charges must be commensurate~\cite{weinberg1995quantum2}. Conversely, if the hypercharges
are commensurate but otherwise free,  gauge anomaly cancellation of a family of SM chiral fermions implies
gravitational anomaly cancellation~\cite{Lohitsiri:2019fuu} within that family.}
on the local part\footnote{The global part is somewhat tricky to study in
such models, not least because we don't know (though some might say we
don't care) what the gauge group is
(see~\cite{Davighi:2019rcd,Davighi:notyet} for
details).} of the anomaly. 
Apart from one crucial
detail (to which we return shortly), the local anomaly depends only on the Lie
algebra which, when the rank increases by one, is unambiguously isomorphic to $su(3) \oplus su(2) \oplus u(1) \oplus
u(1)$.\footnote{When the rank increases by more than one, this is
    a subalgebra and all of the considerations in this work remain
    applicable, but there will, of course, be yet further
    constraints. For a concrete example in which two additional $U(1)$ factors
    are gauged, viz. those corresponding to baryon and lepton numbers, see
    \cite{FileviezPerez:2010gw}.}

{ The main purpose of our work is to study and restrict the
  parameter space of additional-$u(1)$ charges of chiral fermions, thus
  informing the model-building of SM extensions.}

The gauged, spontaneously broken $U(1)$ subgroup  leads to a massive  SM-neutral
spin-1 particle which we may dub a $Z^\prime$.
Models with $Z^\prime$s have been studied exhaustively in the
  literature: 
 to explain dark
  matter~\cite{Okada:2010wd,Nakayama:2011dj,Allanach:2015gkd,Okada:2016tci,Okada:2016gsh,Okada:2017dqs,Agrawal:2018vin,Okada:2018tgy}, the anomalous magnetic moment of the muon~\cite{Heeck:2011wj},
axions~\cite{Berenstein:2010ta} or leptogenesis~\cite{Chen:2011sb},
proton stabilisation~\cite{Carone:1996nd}, supersymmetry breaking~\cite{Kaplan:1999iq},
fermion masses and mixing
(via the Froggatt-Nielsen 
mechanism)~\cite{Froggatt:1978nt}, and, most recently~\cite{Gauld:2013qba,Buras:2013dea,Buras:2013qja,Altmannshofer:2014cfa,Buras:2014yna,Crivellin:2015mga,Crivellin:2015lwa,Sierra:2015fma,Allanach:2015gkd,Crivellin:2015era,Celis:2015ara,Greljo:2015mma,Altmannshofer:2015mqa,Falkowski:2015zwa,Chiang:2016qov,Becirevic:2016zri,Boucenna:2016wpr,Boucenna:2016qad,Ko:2017lzd,Alonso:2017bff,Alonso:2017uky,1674-1137-42-3-033104,CHEN2018420,Faisel:2017glo,PhysRevD.97.115003,Bian:2017xzg,PhysRevD.97.075035,Bhatia:2017tgo,Allanach:2017bta,Allanach:2018odd,Allanach:2018lvl,Duan:2018akc,Geng:2018xzd,Kawamura:2019rth,Allanach:2019mfl,Dwivedi:2019uqd,Allanach:2019iiy,Altmannshofer:2019xda,Calibbi:2019lvs},  apparent 
lepton family non-universality (FNU)
in certain rare neutral current $B$-meson 
decays~\cite{Aaij:2014ora,Aaij:2017vbb,Hiller:2003js}. Several of these
applications require FNU couplings of the $Z^\prime$, corresponding to FNU
charge assignments to its underlying $U(1)$ gauge group.
There
  is also a generic phenomenological motivation for $Z^\prime$ fields
  as follows: the existence of neutrinos and dark matter, among other
  things, motivates the existence of a rich hidden sector, and it is
  natural to wonder whether this hidden sector, just like the visible
  sector, has a gauge symmetry of its own. If that gauge symmetry
  algebra has
  a $u(1)$ factor, then the SM fermions can be charged with respect to
it, giving us a possible `portal' to the hidden sector.

Returning now to our discussion of the algebra $su(3) \oplus su(2) \oplus u(1) \oplus
u(1)$, we can choose a basis ${X,Y}$ for the subalgebra $u(1) \oplus 
u(1)$ in which $Y$ corresponds to the SM hypercharge and $X$ to the
$Z^\prime$, such that the $X$ anomaly
cancellation conditions (ACCs)
become\footnote{Given that the
    symmetry corresponding to the $Z^\prime$ is non-linearly realised,
    there is always the possibility that the anomalies from chiral
    fermions do not
    vanish, but are rather compensated by, {\em e.g.}\/ a
    Wess-Zumino-Witten term. We ignore this possibility here.
}
\begin{subequations}\label{Famonamly}
\begin{align}
3^2 X:\quad & 0 = \sum_{j=1}^3\left(2Q_j+U_j+D_j\right), \label{P1}\\
2^2 X:\quad & 0 =\sum_{j=1}^3\left(3Q_j+L_j\right),\label{P2}
\end{align}
\begin{align}
Y^2 X:\quad & \begin{aligned}0 =\sum_{j=1}^3\left(Q_j\right.+8U_j+2&D_j\\& \left.+3L_j+6E_j\right),\end{aligned}\label{P3}\\
YX^2:\quad & 0 =\sum_{j=1}^3\left(Q_j^2-2U_j^2+D_j^2-L_j^2+E_j^2\right),\label{P4}\\
\text{grav}^2 X:\quad &\begin{aligned} \sum_{i=1}^n
x_i =-\sum_{j=1}^3\left(6Q_j\right.+3&U_j+3D_j\\&\left.+2L_j+E_j\right),\end{aligned}\label{P5}\\
X^3:\quad &\begin{aligned}\sum_{i=1}^n
x_{i}^3 =-\sum_{j=1}^3\left(6Q_j^3\right.+3&U_j^3+3D_j^3 \\ &\left.+2L_j^3+E_j^3\right),\end{aligned}\label{P6}
\end{align}
\end{subequations}
where $F_j$ denote the
charges of SM fermions\footnote{As usual, we consider all
    Weyl fermions as being left-handed.} ($F \in \{
Q,U,D,L,E\}$, $j \in \{1,2,3\}$) with respect to the 
$Z^\prime$ and
$x_i$ denote the charges of $n \geq 0$ hypothetical
SM-singlet 
fermions with respect to the $Z^\prime$.
We posit that SM-singlet fermions are highly likely to exist in Nature given the fact that
neutrinos are massive  and the simplest way of providing neutrino masses is
via Yukawa terms with SM-singlet fermions
(they also have a number of other phenomenological
applications, such as dark matter). By way of shorthand, we
refer to the 
$F_j$ as `visible charges' and the $x_i$ as `invisible charges.'
Most previous studies of these ACCs have made additional assumptions,
  those of
  family universality 
  or two zero-charged families being particularly common.
  We wish our   analysis to 
  be more general, ultimately allowing
the visible charges to freely vary between 
generations. This will lead to a larger set of solutions that contains subsets
with additional assumptions.
Aside from being more general,
family dependence is often desirable in applications to flavour physics.

Now comes the crucial point at which the global structure of the gauge
group plays a r\^{o}le. In the general case, the charges must be
real,
but if the group is compact (as we might expect on the basis of
a variety of theoretical arguments and empirical observations, the most compelling of which are perhaps the
apparent unification of gauge couplings and the fact that the observed
electric
charges themselves appear to be commensurate), then the $X$
charges must be commensurate, and
since (\ref{P1})-(\ref{P6}) are invariant under an overall real rescaling, we
may take 
them to be rational.

Of course, by clearing denominators, we could also take them to be
coprime integers\footnote{That is, integers whose
greatest common divisor is 1.} and indeed we will always write charges in
this way. Thus, for example, we  write the SM fermion's hypercharges as 
$y_{Q_j}=+1,\ y_{U_j}=-4,\ y_{D_j}=+2,\ y_{L_j}=-3,\ y_{E_j}=+6$.
Doing so avoids having to worry about annoying
normalisation factors.\footnote{It also allows us to mention
another subtle point. Depending on the choice of the global gauge
group, not all rational points lead to admissible
representations. Suppose, for example, that the $Z^\prime$ together
with the $W$ and $Z$ gauge bosons actually
make up the group $U(2)$. In this case, the usual global Witten
anomaly of $SU(2)$ is absent and appears instead as a local
anomaly. The number of even-dimensional representations of $U(2)$ must
be even, and if it is non-zero, this implies that the number of odd
integer charges must be even; for details, see~\cite{Davighi:2019rcd,Davighi:notyet}.
We will concentrate here on
the case in which the global group is assumed to be
$SU(3)\times SU(2)\times U(1)\times U(1)$ (which is universal in the
sense that it covers any other compact group with the same algebra),
such that all rational points are allowed.} But considering them to
lie in the field of
rational numbers allows us to benefit from the not inconsiderable
machinery, and geometric insights, of (projective) algebraic geometry.
Indeed, the 6 equations (\ref{Famonamly}) are
homogeneous in $15+n$ unknowns with coefficients in
$\Q$ and so, given any field extension $k$ of $\Q$ (such as $\Q$, $\R$, or
$\C$) they define a projective variety in the $14+n$ dimensional
projective space over $k$, i.e.\ the space of lines through the origin
in the affine space $k^{15+n}$.

The holy grail would be to find all $k$-points for each value of
$n$. For generic polynomial equations, this is hard enough even
for the case $k=\R$ (corresponding to a non-compact gauge group); in
the case $k=\Q$ of interest to us (corresponding
to a compact gauge
group), the best efforts of number theorists over the millennia have
yielded scant reward.\footnote{The state of the art is elliptic
  curves, described by a single cubic equation in 2-d.}

The outlook as regards the search for the holy grail is thus somewhat
bleak.
Fortunately (or unfortunately, depending on one's point of view), 
a general solution
is hardly required, given that so little is known
about SM singlet fermions. It seems likely that they exist (given that
neutrinos are massive), but we don't know how many there are (two or more
suffice to fit neutrino data) and we are certainly not yet in a position
to measure their charges with respect to a $Z^\prime$ boson which is
itself yet to be discovered and may well not exist.
So, at least for the time being, it seems
reasonable to leave the number theorists in peace in their ivory
towers and to focus our attention instead on questions which are of
more immediate interest to phenomenologists. Happily, we will find
that most such questions are also easy enough for phenomenologists to answer. 

Paramount among such questions, in our opinion, is the following:
can we find all possible values of the {\em visible}\/ charges $F_j$ for which there
exists a solution of the ACCs (\ref{Famonamly}) for some $n$ and $x_i$? We may
  wish to know the visible charges because these largely determine the phenomenology
  of the $Z^\prime$.
On the other hand, since
$n$ and $x_i$ are almost completely unconstrained by observations,
we hardly need to know the values they take in particular solutions;  
rather, we might care only whether such values exist.

To this end, our strategy is to first give a general solution of
  the equations, namely (\ref{P1}-\ref{P4}), that only involve the
  visible charges $F_j$. The first 3 of these equations are linear and
are thus trivial to solve over $\Q$. The fourth and last equation is
quadratic and finding its general solution is also a triviality
(albeit an algebraically unpleasant one), once we employ
some insight from geometry. Indeed, suppose that we are somehow able to find
just 1 
rational point. We may then construct all rational lines through
that point. Each such line must either lie in the surface, in which case every
rational point on it gives another solution, or it must intersect the
surface in another rational point,\footnote{This point may, of course, coincide
with the original point -- i.e.\ we have a double point -- or the other
point of intersection may be `at infinity' in affine space.} giving us
a way to generate new solutions from old ones. Moreover, since any two
points (in either affine or projective) space are joined by a line,
all solutions can be obtained in this way.

The 2 equations which remain, (\ref{P5}) and
(\ref{P6}), involve the invisible charges and involve a cubic, so are typically much
harder to solve. Instead we attack the simpler problem of fixing $n$ and asking whether a solution
exists. To this end, it is 
convenient to write (\ref{P5}) and
(\ref{P6}) in the form
\begin{align}
\sum_{i=1}^n
x_i &= J, \label{P7}\\
\sum_{i=1}^n
x_{i}^3 &=M+J^3, \label{P8}
\end{align}
where
\begin{align}J&:=-\sum_{j=1}^3(6Q_j+3U_j+3D_j+2L_j+E_j), \label{Jdef} \\
  M&:=-\sum_{j=1}^3(6Q_j^3+3U_j^3+3D_j^3+2L_j^3+E_j^3)-J^3.
  \label{Mdef}
\end{align}
and evidently $M,J\in \mathbb{Z}$. In the case where two families of
charges are set to zero, we will see that $M=0$, and that this allows
us to solve (\ref{P7}) and (\ref{P8}) generally. The form of $J$ in
this case allows us to connect the solution to this problem to the
solution obtained for the visible charges, allowing the general
solution to the full set of ACCs to be found.

In more general cases, including the apparently very similar
  family-universal (FU)
  case, we are
  not able to obtain a general solution. But in this FU case, we show in \S\ref{sec:famU} that (\ref{P5}) and
(\ref{P6}) always have a solution
over $\Z$ for $n\geq 3$. In the fully family non-universal (FNU) case, we
similarly show in \S\ref{sec:InvisbleACCs} that (\ref{P5}) and
(\ref{P6}) always have a solution for $n \geq 5$. 
Thus, that any of the solutions to (\ref{P1}-\ref{P4})
can be extended to a solution of the full set of ACCs
(\ref{P1}-\ref{P6}) when $n$ is large enough.
In contrast, we shall show that sometimes a solution cannot be found for
$n\leq 2$ in the FU case and $n \leq 4$ in the FNU
case. We catalogue the values of $M$ and $J$ for which solutions can be
found for $n\leq 2$ in the FU case and $n \leq 3$ in the
FNU case.\footnote{There is a small
lacuna in the case $n=4$, which can be traced back to the difficulty of
solving a cubic equation in 1 unknown over $\Z$.} These results hinge
crucially on the non-so-obvious fact that $M \in 6\Z$. For all
cases where a solution exists, we provide a general parameterisation of the
visible charges.  

Thus, we obtain a factorisation of the problem in a way that
ought to be adequate for phenomenologists' needs: a general parametric solution
to the allowed visible charges is given;
for any such solution one can
be sure that a suitable set of invisible charges exists, if $n$ is
large enough.

In an `anomaly-free atlas'~\cite{Allanach:2018vjg}, solutions of the ACCs considered here
  were found numerically for $|F_j|\leq 10$ and $n=0,1,2$ or 3. In the present
  paper, we are
  interested in analytic solutions without any such restrictions upon
  $F_j$.

The outline of our paper is as follows. 
We begin in \S\ref{sec:oneFam} by considering the simple case where
only one family is charged. Here we will show that $M=0$, and as a
consequence the ACCs can be solved exactly, using previously known
results.
In \S\ref{sec:famU}, we
discuss the FU case. In \S\ref{sec:famNonU}
we discuss the more general case where visible charges vary between generations.
We summarise and conclude in \S\ref{sec:conc}.
Though it is something of a curiosity, it turns out that one can also
find a general solution for odd $n$ in the family universal case; this solution is given in Appendix~\ref{sec:noddFU}. 
In Appendix~\ref{sec:catch}, we supply
  additional parameterisations of the solutions for the visible charges
  which avoid the need to consider the 
  degenerate  cases where our main parameterisation yields entire lines
  of solutions. The special case where there are only two
independent families of charges (which may be phenomenologically
relevant given the apparent similarity of the 2 light SM families) is dealt with in Appendix~\ref{sec:semi}. 

\section{One-Family Case} \label{sec:oneFam}
The simplest case we examine is where two families have zero charge and only one family is charged. As such we have $F_2=F_3=0$ and define $F_1=F$. 
We start our analysis by
imposing (\ref{P1}-\ref{P3}), which imply
\begin{align}
D=-2Q-U,\; L=-3Q,\; E=2Q-U. \label{Param}
\end{align}
Substituting this into (\ref{P4}) yields
$0$.\footnote{This has a geometric explanation, and holds for
    generic choices of representations of the five fermionic
      species. (\ref{Param}) defines a line $L$ in
    $P\mathbb{Q}^4$, which must pass through the point corresponding to
    hypercharge 
    assignments. The quadratic volume defined by
    (\ref{P4}) must
    also pass through the hypercharge assignment point
    and so it either must intersect $L$ at one other
    point or $L$ must lie within the quadratic volume itself.
    Since (\ref{P3}) implies
    that 
    the gradient of 
    (\ref{P4}) 
    in the direction of the hypercharge point is zero, $L$ lies within 
    the quadratic volume.  This argument would not work for a larger number
    of fermionic species, since then (\ref{Param}) would define a higher dimensional space, rather then a line.} Thus (\ref{Param}) provides a general 
solution to the visible ACCs (\ref{P1}-\ref{P4}) for any~$Q,U \in \Z$.

Substituting (\ref{Param}) into (\ref{Jdef}) and (\ref{Mdef}) gives $J=4Q+U$
and $M=0$ and so we reduce (\ref{P7})
  and (\ref{P8}) to 
\begin{align}
\sum_{i=1}^n x_i =J, \quad \sum_{i=1}^n x_i^3=J^3.
\end{align}
For $n=0$, there exists a solution to all ACCs iff.\ $J=0$, corresponding to a
one-parameter family of solutions given by (\ref{Param}) with $U=-4Q$ already
found in Ref.~\cite{Allanach:2018vjg}.
For $n>0$,
by defining $x_{n+1}=-J$, we obtain a set of diophantine equations equivalent
to the ACCs of a pure $U(1)$ gauge theory, the general solution to which is
known~\cite{Costa_Dobrescu_Fox_2019,Allanach:2019gwp}, and which parameterise
$x_i$ in terms of a set of parameters $R$ (say). Thus in this case the general
solution to the ACCs can be found in terms of the parameters $\{R,Q\}$,
\begin{align}
U&=-4Q-x_{n+1}(R),\; D=2 Q+x_{n+1}(R),\; L=-3 Q,\nonumber \\ E&=6 Q+x_{n+1}(R),\;\; x_i=x_i(R)\ \forall\ 1\le i\le n.
\end{align}

\section{Family Universal Case} \label{sec:famU}
The next step in our discussion is the FU case.  
Like for the one family case above, we start our analysis by
imposing (\ref{P1}-\ref{P3}), leading to (\ref{Param}) again, but where $F_1=F_2=F_3=F$.\footnote{This corresponds to the well-known (see
    {\em e.g.} \cite{weinberg1995quantum2}) fact
  that, given the $SU(3)\times SU(2)$ representation content of the SM
fermions, the only gaugeable, FU $U(1)$ charges are a linear combination
of the usual hypercharge and $B-L$.}

Substituting (\ref{Param}) into (\ref{Jdef}), (\ref{Mdef})
gives us $J=3(4Q+U)$ and $M=-24(4Q+U)^3$. Notice, here is where our analysis
diverges from the one-family case. The fact that $M\ne 0$ prevents us from
solving (\ref{P7}) and (\ref{P8}) in a way identical
to~\cite{Costa_Dobrescu_Fox_2019,Allanach:2019gwp}.  As such we are left to
study (\ref{P7}) and (\ref{P8}) on a case-by-case basis, looking at the
existence of 
solutions. For $n=0$, it is obvious that we need $J=0\Rightarrow M=0$, leading
to the same one-parameter set of solutions as in the $n=0$ one-family case. For $n=1$,
$x_1=J$ and $x_1^3=M+J^3$, so $M=0\Rightarrow J=0$ as well. For $n=2$, $x_1+x_2=J$ and
$x_1^3+x_2^3=M+J^3$, eliminating $x_2$ from the latter equation gives  
\begin{align}
9K x_1^2-27 K^2 x_1+24 K^3=0,
\end{align}
where $K\equiv J/3\in \mathbb{Z}$ and we have used that $M=-24K^3$.
This equation has no real roots (ergo no integer roots) unless $K=0\Rightarrow
J= M=0$ again.  
For $n\ge 3$, we always have a solution given by $x_1=x_2=x_3=K=(4Q+U)$ and
$x_i=0$ for $i> 3$. While this is {\em a}\/ solution, we do not claim that it
is the {\em most general}\/ one. 
However, in the case of $n$-odd it is an oddity that we can find the most
general solution.\footnote{It arises because the cubic hypersurface
  has double points, and so all solutions can be obtained by
  constructing lines through such a point.} This is detailed in Appendix~\ref{sec:noddFU}.

We will see in \S\ref{sec:InvisbleACCs}, that we will be less lucky when
considering the equivalent problem in the FNU case, and we will have to resort
to less trivial techniques. 

\section{Family Non-Universal Case} \label{sec:famNonU}
We now move to the FNU case and find ourselves in the fortunate situation that the ACCs
(\ref{P1}-\ref{P4}) (those
that depend only on visible matter) can, when considered alone, be
solved generally using straightforward, if unpleasant, techniques from diophantine
analysis. To do so, 
it helps to apply a common $GL(3,\mathbb{Z})$ transformation
  to $\{F_1,F_2,F_3\}$ similar to that in Ref.~\cite{Allanach:2018vjg}, so as to recast the equations into
  a simpler form, whilst remaining within the realm of the
  integers. 

To wit, we set $F_+=F_1+F_2+F_3$,
  $F_\alpha=F_1-F_2$, and $F_\beta = F_2 + F_3$. 
The
  judiciousness of this transformation is twofold.  Firstly, the
 linear equations (\ref{P1}-\ref{P3}) become dependent on $F_+$ only, whose general solution over $\mathbb{Z}$ is immediately seen to be 
\begin{align} \label{eq:plusolu}
\begin{split}
D_+=-2Q_+-U_+, &\quad L_+=-3Q_+, \\E_+=2&Q_+-U_+,
\end{split}
 \end{align}
 written in term of two arbitrary parameters $Q_+,U_+\in\mathbb{Z}$, which we will see are further constrained by the quadratic. Secondly, it makes it easy for us to
 recover the results in situations with charge universality between any
 two families, which without loss of generality corresponds to
 setting $F_\alpha =0 $; this is considered in Appendix~\ref{sec:semi}.
 
In the new variables the quadratic equation becomes
 \begin{equation}
  X^T HX=0 \label{eq:quad}
\end{equation}
where we have used (\ref{eq:plusolu}) to replace $D_+$, $L_+$ and $E_+$.
It is a homogenous diophantine equation of
degree $2$ in the entries of the 12-tuple
\begin{align}
\begin{split}
X := (Q_+,U_+,Q_\alpha,Q_\beta,U_\alpha,U_\beta,&D_\alpha,D_\beta,\\ &L_\alpha,L_\beta,E_\alpha,E_\beta).
\end{split}
\end{align}
$H$ is a $12\times 12$
symmetric matrix with integer entries,
the upper right triangle of which is
 \begin{align}
   &H =\nonumber \\&{\scriptsize \left( \begin{array}{rrrrrrrrrrrr}
  0 & 0 &-2 &-4 & 0 & 0 & 4 & 8 &-6 & -12 &-4 &-8 \\
    & 0 & 0 & 0 & 4 & 8 & 2 & 4 & 0 & 0 & 2 & 4 \\
    &   & 2 & 3 & 0 & 0 & 0 & 0 & 0 & 0 & 0 & 0 \\
    &   &   & 6 & 0 & 0 & 0 & 0 & 0 & 0 & 0 & 0 \\
    &   &   &   &-4 & -6 & 0 & 0 & 0 & 0 & 0 & 0 \\
    &   &   &   &   &-12& 0 & 0 & 0 & 0 & 0 & 0 \\
    &   &   &   &   &   & 2 & 3 & 0 & 0 & 0 & 0 \\
    &   &   &   &   &   &   & 6 & 0 & 0 & 0 & 0 \\
    &   &   &   &   &   &   &   &-2 & -3 & 0 & 0 \\
    &   &   &   &   &   &   &   &   & -6 & 0 & 0 \\
    &   &   &   &   &   &   &   &   &   & 2 & 3 \\
    &   &   &   &   &   &   &   &   &   &   & 6 \\    
    \end{array} \right). \label{Hmat}}
  \end{align}
We shall now show that (\ref{eq:quad}), which
defines a hypersurface $\Gamma \subset \mathrm{P}\mathbb{Q}^{11}$,
can be solved generally over $\mathbb{Z}$ given only one non-trivial 
solution. We consider lines, $L=\alpha \tilde X+\beta R$ through a known solution $\tilde X\in \mathrm{P}\mathbb{Q}^{11}$, where $R\in \mathrm{P}\mathbb{Q}^{11}$, and  $[\alpha:\beta]\in \mathrm{P}\mathbb{Q}^{1}$. On substitution into (\ref{eq:quad}) we obtain
\begin{align}\label{alphabetacondi}
\beta(2 R^T H \tilde X\alpha+R^T H R \beta)=0.
\end{align} 
Thus either the line $L$ intersects $\Gamma$ at $\beta=0$ (returning the point $\tilde X$) and at $[\alpha:\beta]=[R^THR:-2R^TH\tilde X]$, or $L\subset \Gamma$  with (\ref{alphabetacondi}) being automatically satisfied for any $\alpha$ and $\beta$.

For $L\nsubseteq \Gamma$, after returning to affine space, the solutions generated from the second intersection point are 
\begin{align}
X=\frac{k}{\mathrm{GCD}(X^\prime)} &X^\prime \nonumber \\\text{~where~} &X^\prime=(R^T H R) \tilde X-2 (R^TH \tilde X)R,\label{genSol}
\end{align}
where $k\in \mathbb{Z}$ is an overall factor,
and $\mathrm{GCD}(X^\prime)$ denotes the greatest common divisor of the
integers in $X^\prime$.

\subsection{Solution for SM chiral fermion charges}
A specific choice of $\tilde X$ has $Q_\alpha=1$ and $L_\alpha=1$, with all
other parameters zero.  Using the formula~(\ref{genSol}) we obtain
\begin{align}
 Q_+^\prime=2R_{Q_+}\Lambda,\  U_+^\prime=2R_{U_+}\Lambda,\ Q_\alpha^\prime=2R_{Q_\alpha} \Lambda+\Sigma,\nonumber\\ 
  Q_\beta^\prime=2R_{Q_\beta} \Lambda,\  U_\alpha^\prime=2R_{U_\alpha} \Lambda,\  U_\beta^\prime=2R_{U_\beta} \Lambda \nonumber\\
  D_\alpha^\prime=2R_{D_\alpha} \Lambda,\  D_\beta^\prime=2R_{D_\beta}
  \Lambda,\  L_\alpha^\prime=2R_{L_\alpha}
  \Lambda+\Sigma,\nonumber\\ 
   L_\beta^\prime=2R_{L_\beta} \Lambda,
  E_\alpha^\prime=2R_{E_\alpha} \Lambda,\  E_\beta^\prime=2R_{E_\beta}
  \Lambda, \label{eq:eg}
\end{align}
where
\begin{align}
\Lambda &=(8R_{Q_+}+2R_{L_\alpha}+3
R_{L_\beta}-2R_{Q_\alpha}-3R_{Q_\beta}), \label{eq:Lambda} \\
\Sigma&=R^THR. 
\label{eq:Sigma}
\end{align}
We invert the $GL(3,\mathbb{Z})$ transformation used above 
using (\ref{eq:plusolu}) and (\ref{eq:eg}) to
yield a parameterisation of a
solution for the visible charges
\begin{align}
Q_1&=2\Lambda(R_{Q_+}-R_{Q_\beta}),\nonumber \\Q_2&=
2\Lambda(R_{Q_+}-R_{Q_\alpha}-R_{Q_\beta})-\Sigma,\nonumber
\\Q_3&=2\Lambda(2R_{Q_\beta}+R_{Q_\alpha}-R_{Q_+})+\Sigma \nonumber \\
U_1&=2\Lambda(R_{U_+}-R_{U_\beta}),\nonumber \\U_2&=
2\Lambda(R_{U_+}-R_{U_\alpha}-R_{U_\beta}),\nonumber
\\U_3&=2\Lambda(2R_{U_\beta}+R_{U_\alpha}-R_{U_+}) \nonumber\\
D_1&=-2\Lambda(2R_{Q_+}+R_{U_+}+R_{D_\beta}),\nonumber \\D_2&= -2\Lambda(R_{D_{\alpha }}+R_{D_{\beta }}+2 R_{Q_+}+R_{U_+}),\nonumber \\D_3&=2 \Lambda  \left(R_{D_{\alpha }}+2 R_{D_{\beta }}+2 R_{Q_+}+R_{U_+}\right)\nonumber\\
L_1&=-2 \Lambda  \left(R_{L_{\beta }}+3 R_{Q_+}\right),\nonumber
\\L_2&=-2\Lambda  \left( R_{L_{\alpha }}+ R_{L_{\beta }}+3
R_{Q_+}\right)-\Sigma,\nonumber \\L_3&=2\Lambda  \left( R_{L_{\alpha }}+2 R_{L_{\beta }}+3 R_{Q_+}\right)+\Sigma \nonumber\\
E_1&=2 \Lambda  \left(2 R_{Q_+}-R_{E_{\beta }}-R_{U_+}\right),\nonumber \\E_2&=2 \Lambda
\left(2 R_{Q_+}-R_{E_{\alpha }}-R_{E_{\beta
}}-R_{U_+}\right),\nonumber \\E_3&=2 \Lambda  \left(-2 R_{Q_+}+R_{E_{\alpha
}}+2 R_{E_{\beta }}+R_{U_+}\right) \label{eq:param}
\end{align}
in terms of the 12 integer valued variables
\begin{align*}\{R_{Q_+},R_{U_+},R_{Q_\alpha},R_{Q_\beta},R_{U_\alpha},&R_{U_\beta},R_{D_\alpha},R_{D_\beta},\\ &R_{L_\alpha},R_{L_\beta},R_{E_\alpha},R_{E_\beta}\}.\end{align*}

Note that for this $\tilde X$, those solutions for which $Q_\alpha=L_\alpha$
lie on lines which themselves lie in $\Gamma$. These points can be caught by
either scanning over all $R$ and noting that when $R^T H\tilde X=0$ all
rational points on the line are solutions, or by taking a series of different
$\tilde X$. In Appendix~\ref{sec:catch} we give a set of $\tilde X$, such that
all solutions can be found as the unique second intersection point of $\Gamma$
with a line through $\tilde X$.
\subsection{Gauge Invariant Yukawa Couplings} \label{sec:gauge}
As an example of how the parameterisation in (\ref{eq:param}) could be used,
we ask the question: 
what are conditions on $R$ from the existence of Yukawa couplings? 

First, we note that a necessary and sufficient condition for the
presence of {\em all} possible gauge invariant Yukawa couplings is
that 
we have family universality, as shown in Ref.~\cite{Allanach:2018vjg}. We analysed this
case in \S~\ref{sec:famU}. 
Going to the family non-universal case, 
we
write the Yukawa couplings in the Lagrangian density, 
\begin{align} \label{eq:yukawaterm}
{\mathcal L}_Y=(Y_U)_{ij} q_iH^c u_j^c+(Y_D)_{ij} q_i H d_j^c+(Y_E)_{ij} l_i H e_j^c + \mathrm{h.c.},
\end{align}
where $\{q_i,u_i^c,d_i^c,l_i,e_i^c\}$ are the left-handed fermion fields in
the family indexed by $i$, $H$ is the SM Higgs doublet field and
$(Y_{U,D,E})_{ij}$ are 3 by 3 matrices of dimensionless Yukawa couplings for
the up quarks, down quarks and charged leptons, respectively.
In the SM, the top Yukawa coupling is $\sim {\mathcal{O}( 1)}$, so we require that it is
allowed by the additional $U(1)$ symmetry at the unbroken level, i.e.\
\begin{align}
H_0=Q_1+U_1, \label{basic}
\end{align}
where 
$H$ a charge of $H_0$ with respect to the $Z^\prime$.
Identifying the left and (anti-)right-handed top fields with $Q_1$ and
$u_1^c$, we then obtain from (\ref{eq:param}) the charges for field operators
multiplying the Yukawa coupling in Table~\ref{tab:charges}. If the charge
combination is non-zero, then the Yukawa coupling is absent from the 
unbroken theory (by gauge symmetry).
\begin{center}
\begin{table*}[t]
\begin{tabular}{cc}\hline
$ij$ & charge \\ \hline 
\multicolumn{2}{c}{$(Y_U)$} \\
11 & $0 $ \\ 
12 & $-2 \Lambda  R_{U_\alpha } $ \\ 
13 & $2 \Lambda  \left(R_{U_\alpha }+3 R_{U_\beta }-2 R_{U_+}\right) $ \\ 
21 & $-2 \Lambda  R_{Q_\alpha }-\Sigma $ \\ 
22 & $-2 \Lambda  \left(R_{Q_\alpha }+R_{U_\alpha }\right)-\Sigma $ \\ 
23 & $2 \Lambda  \left(-R_{Q_\alpha }+R_{U_\alpha }+3 R_{U_\beta }-2 R_{U_+}\right)-\Sigma $ \\ 
31 & $2 \Lambda  \left(R_{Q_\alpha }+3 R_{Q_\beta }-2 R_{Q_+}\right)+\Sigma $ \\ 
32 & $2 \Lambda  \left(R_{Q_\alpha }+3 R_{Q_\beta }-2 R_{Q_+}-R_{U_\alpha }\right)+\Sigma $ \\ 
33 & $2 \Lambda  \left(R_{Q_\alpha }+3 R_{Q_\beta }-2 R_{Q_+}+R_{U_\alpha }+3 R_{U_\beta }-2 R_{U_+}\right)+\Sigma $ \\  \hline
\multicolumn{2}{c}{$(Y_D)$} \\
11 & $-2 \Lambda  \left(R_{D_\beta }+2 R_{Q_\beta }+R_{U_\beta }\right) $ \\ 
12 & $-2 \Lambda  \left(2 R_{D_\alpha }+R_{D_\beta }+2 R_{Q_\beta }-R_{Q_+}+R_{U_\beta }\right) $ \\ 
13 & $2 \Lambda  \left(R_{D_\alpha }+2 \left(R_{D_\beta }-R_{Q_\beta }+2 R_{Q_+}+R_{U_+}\right)-R_{U_\beta }\right) $ \\ 
21 & $-2 \Lambda  \left(R_{D_\beta }+R_{Q_\alpha }+2 R_{Q_\beta }+R_{U_\beta }\right)-\Sigma $ \\ 
22 & $-2 \Lambda  \left(2 R_{D_\alpha }+R_{D_\beta }+R_{Q_\alpha }+2 R_{Q_\beta }-R_{Q_+}+R_{U_\beta }\right)-\Sigma $ \\ 
23 & $2 \Lambda  \left(R_{D_\alpha }+2 R_{D_\beta }-R_{Q_\alpha }-2 R_{Q_\beta }+4 R_{Q_+}-R_{U_\beta }+2 R_{U_+}\right)-\Sigma $ \\ 
31 & $ -2 \Lambda  \left(R_{D_\beta }-R_{Q_\alpha }-R_{Q_\beta }+2 R_{Q_+}+R_{U_\beta }\right)+\Sigma $ \\ 
32 & $ -2 \Lambda  \left(2 R_{D_\alpha }+R_{D_\beta }-R_{Q_\alpha }-R_{Q_\beta }+R_{Q_+}+R_{U_\beta }\right) +\Sigma$ \\ 
33 & $2 \Lambda  \left(R_{D_\alpha }+2 R_{D_\beta }+R_{Q_\alpha }+R_{Q_\beta }+2 R_{Q_+}-R_{U_\beta }+2 R_{U_+}\right)+\Sigma $ \\  \hline
\multicolumn{2}{c}{$(Y_E)$} \\
11 & $-2 \Lambda  \left(R_{E_\beta }+R_{L_\beta }+R_{Q_\beta }+R_{U_\beta }\right) $ \\ 
12 & $-2 \Lambda  \left(R_{E_\alpha }+R_{E_\beta }+R_{L_\beta }+R_{Q_\beta }+R_{U_\beta }\right) $ \\ 
13 & $-2 \Lambda  \left(-R_{E_\alpha }-2 R_{E_\beta }+R_{L_\beta }+R_{Q_\beta }+4 R_{Q_+}+R_{U_\beta }-2 R_{U_+}\right) $ \\ 
21 & $-2 \Lambda  \left(R_{E_\beta }+R_{L_\alpha }+2 R_{L_\beta }+R_{Q_\beta }+R_{U_\beta }\right)-\Sigma $ \\ 
22 & $-2 \Lambda  \left(R_{E_\alpha }+R_{E_\beta }+R_{L_\alpha }+2 R_{L_\beta }+R_{Q_\beta }+R_{U_\beta }\right)-\Sigma $ \\ 
23 & $-2 \Lambda  \left(-R_{E_\alpha }-2   R_{E_\beta }+R_{L_\alpha }+2 R_{L_\beta }+R_{Q_\beta }+4 R_{Q_+}+R_{U_\beta }-2 R_{U_+}\right)-\Sigma $ \\ 
31 & $ -2 \Lambda  \left(R_{E_\beta }+R_{L_\alpha }+R_{L_\beta }+R_{Q_\beta }+R_{U_\beta }\right) +\Sigma$ \\ 
32 & $ -2 \Lambda  \left(R_{E_\alpha }+R_{E_\beta }+R_{L_\alpha }+R_{L_\beta }+R_{Q_\beta }+R_{U_\beta }\right)+\Sigma $ \\ 
33 & $2 \Lambda  R_{E_\alpha }+4 \Lambda  R_{E_\beta }-2 \Lambda  \left(R_{L_\alpha }+R_{L_\beta }+R_{Q_\beta }+4 R_{Q_+}+R_{U_\beta }-2 R_{U_+}\right)+\Sigma $ \\  \hline
\end{tabular}
\caption{$Z^\prime$ charges of field operators multiplying each entry of a Yukawa
  matrix from (\ref{eq:param}). In the unbroken $U(1)$ theory, the Yukawa term may only be present
  if the charge is equal to zero. The first column labels the entry
  of the matrix, the second gives the charge. $\Lambda$ and $\Sigma$ are
  given in (\protect\ref{eq:Lambda}) and (\protect\ref{eq:Sigma}),
  respectively. \label{tab:charges}} 
\end{table*}
\end{center}
We note that there are many different choices that one could make, by
identifying the top quark with $Q_{2,3}$ and/or $u_{2,3}$ fields, or indeed by
choosing one of the other sets of solutions shown in
Table~\ref{setTable}. However, rather than provide an encyclopedia of Yukawa couplings for all
possible solutions, our
purpose here is to illustrate how our solutions may be used and what some of
the issues are. 

Since the top, bottom and tau have large masses compared to the other
fermions, we might expect each of 
their Yukawa couplings to be present, being gauge invariant under the
additional $U(1)$ subgroup. 
This subgroup may well ban other tree-level, renormalisable Yukawa couplings, but
because the additional $U(1)$ symmetry is spontaneously broken, we expect small corrections
to any zeroes in the Yukawa matrices. This is the  philosophy behind many
models of fermion masses, including the Third
Family Hypercharge model (TFHM)~\cite{Allanach:2018lvl}, where such hierarchies are
successfully employed to qualitatively explain two features: small quark mixing and the relative
heaviness of the third family of fermions. 
A discussion of neutrino masses
requires an analysis involving the SM-singlets, whose charges we do not yet have a
general solution for, and so we leave these for future work. 

Let us, for the sake of illustration, further identify the tau lepton with
$L_1$ and $e_1$ and the bottom quark with $Q_1$ and $d_1$. 
By consulting Table~\ref{tab:charges}, 
we see that for the (11) entry of each Yukawa matrix to be allowed under the
additional $U(1)$ subgroup,
\begin{align}
H_0=-Q_1-D_1=-L_1-E_1, \label{others}
\end{align}
in addition to (\ref{basic}).
Using  (\ref{eq:param}) the additional constraints become requirements on $R$,
given by either
\begin{align}
R_{U_+}&=-R_{D_\beta}+R_{E_\beta}+R_{L_\beta}-R_{Q_\beta} -R_{Q_+}\nonumber \\
R_{U_\beta}&=-R_{D_\beta}-2R_{Q_\beta},\label{cons}
\end{align}
or the trivial solution $\Lambda=0$, which we discount as being uninteresting,
having only zero $Z^\prime$ charges for the SM fermions.
For the non-trivial solution, (\ref{cons}) reduces the dimensionality of the solutions
from 12 to 10. 

\subsection{A lemma on $M$} \label{sec:Ilemma}
Here we show that $M\in 6\Z$ (see (\ref{Mdef}) for the definition of $M$), a result that will be of use later. From (\ref{P1}) we have $\sum_j^3(U_j+D_j)=0\mod 2$ and from (\ref{P2})
we have $\sum_j^3 L_j=0\mod 3$ which also imply that
$\sum_j^3(U_j^3+D_j^3)=0\mod
2$ and $\sum_j^3 L_j^3=0\mod 3$, respectively. Hence $J=-\sum_j^3E_j\mod 6$ and $M+J^3=-\sum_j^3 E_j^3
\mod 6$, which gives us
$M=3(E_1+E_2)(E_2+E_3)(E_3+E_1)\mod
6$. Since $(E_1+E_2)(E_2+E_3)(E_3+E_1)$ is
always even for any integers $E_i$, we have $M=0\mod 6$ QED\@. Given this it is convenient for us to define $6P=M$. 
\subsection{The number of SM singlets}\label{sec:InvisbleACCs}
Although we were lucky with the visible ACCs, allowing us to solve them
generally, we will be less lucky with the remaining two ACCs
(\ref{P7}-\ref{P8}) which describe the invisible
charges.
As such, we will proceed in a similar manor to the FU case and establish the following facts: for $n\leq 4$ a
solution does not always exist; for $n\ge 5$ a solution always
exists. Furthermore, for $n\leq 3$ we give a complete characterisation
of the solutions. Let us begin with discussion of these cases.
\subsection*{$n \leq 3$}
For $n=0$, the ACCs clearly have a solution iff.\ $P=J=0$. For
any $n \geq 1$, we can eliminate $x_n$  from (\ref{P7}-\ref{P8});
the resulting equation, which must be symmetric under
permutations of the remaining $n-1$ charges, may be recast in terms
of the elementary symmetric polynomials $e_j(x_1,\cdots,x_{n-1})\equiv\sum_{1\le i_1<i_2<\cdots< i_j\le {n-1}}x_{i_1}x_{i_2}\cdots x_{i_j}$ as
\begin{gather}
e_3=2P + e_1 e_2 + Je_1(J-e_1).
\end{gather}
This condition
makes it relatively easy to deduce when we have a solution for $n \leq
3$, as follows.

For $n=1$, we have $e_1=e_2=e_3 = 0 \implies P=0 $, in which case the solution is $x_1 = J$. 

For $n=2$, we have $e_2=e_3=0
\implies 0 = 2P + Je_1(J-e_1)$. This quadratic equation in $e_1$ has a
solution in the integers only if $J$ divides $2P$ (barring the trivial
case $J=0$, for which the solution is $x_1=-x_2$). 
If so, we have the equation 
$e_1^2 - Je_1 -2P/J =0$ with integer coefficients. Since the leading
coefficient is unity, any rational solution is also an integer
solution. To get a rational solution, the discriminant $J^2+8P/J$ must
be square. So we have an integer solution iff.\ $J$ divides $2P$ and $J^2+8P/J$ is square. 
(As a check, our result for $n=2$ subsumes the result for $n=1$, since
if $P=0$, then $J^2+8P/J = J^2$ is certainly square.) The 2 invisible
charges $x_1$ and $x_2$ are then the 2 roots of $x^2 - Jx -2P/J =0$ and if a
solution exists, it is unique up to permutation of the charges. 

For $n=3$, we have $e_3 = 0 \implies 2P + e_1 e_2 + Je_1(J-e_1) =
0$, such that the two charges $x_1$ and $x_2$ are fixed to be the 2
roots of the quadratic equation
\begin{gather}
0= x^2 - e_1 x + e_1J-2P/e_1 - J^2,
\end{gather}
where the third charge is given by $J-e_1$. Since both roots $x_1$ and $x_2$
correspond to
charges, they must both be valued in the integers and so must their product, which
equals $e_1J-2P/e_1 - J^2$. Thus $e_1$ must divide $2P$. As for $n=2$, any
rational solution must then be an integer solution because the leading
coefficient is unity, and a rational solution is obtained iff.\ the
discriminant $(2J-e_1)^2+8P/e_1$ is square.  {\em In toto}, we have that there
exists a solution iff.\ there exists a divisor $e_1$ of $2P$ such that $(2J-e_1)^2+8P/e_1$ is square.
The number of possible solutions is
finite (at least for $P \neq 0$), being at
most (up to permutation of roots) given by the number of divisors of
$2P$.

To recover the result for $n=2$, set $e_1=J$, such that the third root
is $0$. Then we have a solution
iff. $J^2 + 8P/J$ is square, such that $J$ divides $2P$, which are
precisely the conditions found for $n=2$.
\subsection*{$n = 4$}
When $n=4$, we can use similar considerations to those for $n \leq 3$
to conclude that the 3 charges $x_1, x_2, x_3$ should be the 3 solutions of the
cubic
\begin{multline}
x^3 - e_1 x^2 + e_2 x - e_3 = \\x^3 - e_1 x^2 + e_2 x - (2P + e_1 e_2 +
Je_1(J-e_1)) = 0.
\end{multline}
A necessary (but not sufficient) condition for the charges to be integers is that the discriminant
\begin{multline}
-\left(e_1 J \left(J-e_1\right)+2 P\right) \left(27 e_1 J^2-27 e_1^2 J+4 e_1^3+54 P\right)\\-4e_2^3-8 e_2^2e_1^2-4 e_2 e_1 \left(9 e_1 J^2-9 e_1^2 J+e_1^3+18 P\right)
\end{multline}
must be expressible in the form $r^2s^2(r-s)^2,$ for some $r,s \in \Z$. Indeed the
discriminant is defined to be the
product of the squares of the differences of the roots and we are free
to shift the roots by an integer such that one root vanishes, without
changing the discriminant's value. But this condition is difficult to
express in terms of conditions on $P$ and $J$. Thus we content
ourselves with showing that there exist $P$ and $J$ for which no
solution can be found. 

 By explicit evaluation of the
various possibilities for the various charges modulo 9, we find the following conditions on $P$ and
$J$ $(\mathrm{mod}\ 9)$ for a solution to exist:
\begin{align}
\begin{split}
J\ \mathrm{mod}\ 9\in \{1,4,7\}&\text{~requires~}  P\ \mathrm{mod}\ 9\notin\{2,5\},\\
J\ \mathrm{mod}\ 9\in \{2,5,8\}&\text{~requires~} P\ \mathrm{mod}\ 9\notin\{4,7\}.
\end{split}
\end{align}
Any of the 12 cases ($\mathrm{mod}\ 9$) not covered above could furnish us with a counterexample, if
  we could find corresponding values for the visible charges.
A particularly simple counterexample is given by
\begin{align}
Q_1=-2,\ Q_2=0,\ Q_3=2,\nonumber\\ E_1=-1,\ E_2=1,\ E_3=1,\nonumber\\ U_1=-2,\ U_2=0,\ U_3=1,\nonumber \\
 L_1=-1,\ L_2=-1,\ L_3=2,\nonumber\\ D_1=-1,\ D_2=0,\ D_3=2,
\end{align}
which satisfy the ACCs (\ref{P1}-\ref{P4}) but give $(P,J)=(2,1)$.
Thus we see that it is not always possible
to find suitable charges for 4 invisible particles such that all
anomalies cancel. 
\subsection*{$n \geq 5$}
For the case of $n\ge 5$ and for any given $P$ and $J$ there is always
a set of invisible charges which satisfy (\ref{P7}) and (\ref{P8}). To show this, we show it is true for the $n=5$ case, with the $n>5$ cases immediately following by setting the extra charges to zero.

For $n=5$, we set $x_5=J$ and $x_4=-x_1-x_2-x_3$, which immediately
satisfies (\ref{P7}) and reduces (\ref{P8}) to 
\begin{align}\label{n5product}
(x_2+x_3)(x_3+x_1)(x_1+x_2)=-2P.
\end{align}
The choice of charges $x_1=P+1$, $x_2=-P$ and $x_3=-P$ satisfy
the above equation, and, are integer. Thus the integer set of charges
\begin{align}
\left\{P +1,P-1, -P,-P , J\right\}
\end{align}
satisfy (\ref{P7}) and (\ref{P8}) for $n=5$. 

The results of this section are summarised in Table~\ref{invisibleResults}. 
\renewcommand{\arraystretch}{1.5} 
\begin{table*}[t]
\begin{center}
{\small
 \begin{tabular}{|c| >{\centering\arraybackslash}p{2.4in}| >{\centering\arraybackslash}p{3.3in}|}
\hline
$n$ & Solutions exist & Invisible charges are roots $x$ of\\ \hline
0& iff. $ J=0$ and $P=0$ & --\\ \hline
1& iff. $ P=0$ & $x=J$\\ \hline
2& iff. $ J | 2P$ and $J^2+8P/J$ is square & $x^2 - Jx -2P/J =0$\\ \hline
3&iff. $ \exists e_1 $ s.t.\  $e_1|2P$ and $(2J-e_1)^2+8P/e_1$ is square & $(x-J+e_1)(x^2 - e_1 x + e_1J-2P/e_1 - J^2)=0$ \\ \hline
4& only if $ (P,J) \notin \{(\{2,5\},\{1,4,7\}), (\{4,7\},\{2,5,8\})\}$ & $(x-J+e_1) (x^3 - e_1 x^2 + e_2 x - 2P - e_1 e_2 -
Je_1(J-e_1)) = 0$.\\ \hline
$n \ge 5$& $\forall \, P,J$ & \emph{e.g.} $x^{n-5}(x-J)(x-1-P)(x-P+1)(x+P)^2=0$\\ \hline
  \end{tabular} }

\caption{Summary of the solutions to (\ref{P7}) and (\ref{P8}) for various $n$
  in the NFU case. } 
\label{invisibleResults}
\end{center}
\end{table*}
\renewcommand{\arraystretch}{1} 

\section{Summary and Conclusion \label{sec:conc}}

Countless gauge extensions of the SM have been constructed,
most of them\footnote{One notable exception is $SU(5)$, which has been
  studied {\em ad nauseam}\/ in the literature.}
 including a $U(1)$ subgroup in the extension.
In this paper, we have studied 
  possible values of the charges of chiral fermions
  under the additional $U(1)$ with two
  assumptions: (i) that local anomalies cancel and (ii) that the chiral fermion
  content is that of the SM plus a number $n$ of
  SM-singlet chiral fermions. The ACCs split into two
  classes: the first class, with four equations, involves no SM-singlet
  fermionic charges. The second class, with two equations, does involve them. 
  Assumption (ii) is crucial; the assumption of additional SM-singlets allows a general solution to
  the problem, such as it is specified, to be found: the 
  charges of SM chiral fermions under
  the additional $U(1)$ subgroup are parameterised in (\ref{eq:param}). 
  Solving for the SM-singlet charges in general from the second class 
  proves to be much trickier.
  We thus resort to calculating how many of them are
  required to provide a solution to all of the ACCs.

Before tackling the most general problem, we examine two simpler
  and well-motivated cases where only one family is charged and where all visible charges are family
  universal. In the former case we were able to solve the problem in full
  generality, specifying a parameterisation of the solution for the visible
  and SM-singlet charges (of a specified number). 
  For this case, a solution to the first class of ACCs can  always be extended to a
  full solution of all six ACCs when
  there are 3 or more SM singlets. 

We then moved onto the full family non-universal case, solving the first class of ACCs in full
generality, cf (\ref{eq:param}). 
The result is a 12-dimensional parameterization of SM fermion charges, shown
in (\ref{eq:param}). We have illustrated how the parameterization may be used
in practice by examining constraints leading to the presence or absence of
Yukawa couplings under the full gauge invariance  in 
\S~\ref{sec:gauge}.
Requiring gauge invariant third-family Yukawa
couplings restricts the dimensionality of the parameterization further,
and could be used as a basis for more detailed fermion mass model building. 

Our progress for the invisible charges followed a similar path
to the family universal case. We found conditions on functions of $U(1)$
charges of SM fermions
$J$ and $M$, such that a solution to
the first class of ACCs could be extended to the full ACCs, for up to $n=3$
SM-singlet charges. We also showed that the minimum number of SM-singlet charges
such that a solution can always be extended is $5$. We give a solution for
these 5 singlet charges.

If the SM is extended by some non-abelian group $G$ with a $U(1)$ gauge subgroup,
our analysis obviously still applies to the charges of chiral fermions under the $U(1)$
subgroup. One may take our solution parameterisations and then
apply any further
constraints implied by the rest of $G$.

\section*{Acknowledgements}
We thank 
other members of the Cambridge Pheno Working Group for
discussions. 
This work has been partially supported by STFC consolidated grants
ST/P000681/1 and 
ST/S505316/1. BG is also supported by King's College, Cambridge.
\appendix
\section{Solution for invisible charges for the family universal case and $n$-odd\label{sec:noddFU}}
In \S\ref{sec:famU} we had the equations,
\begin{align}
\sum^n_{i=1} x_i&=J=3K\label{ap1}\\
\sum^n_{i=1} x_i^3&=J^3+M=3K^3.\label{ap2}
\end{align}
where $K=4Q+U$. It turns out that (\ref{ap1})-(\ref{ap2}) can be solved exactly for odd
$n$. They imply that
\begin{align}
9\sum^n_{i=1} x_i^3-\left(\sum^n_{i=1} x_i\right)^3=0,
\end{align}
defining a cubic surface $C$ in the projective space $\mathrm{P}\mathbb{Q}^{n-1}$. This cubic surface has a double point (a point on the surface where all derivatives of the surface vanish) at (for $n\ge 3$ and odd)
\begin{multline}
S:=[1:1:1:-1:1:-1:1:\\\cdots:-1:1]\in \mathrm{P}\mathbb{Q}^{n-1}.
\end{multline} 
Consider a line through this point, $L=\alpha S+\beta R$, for
$[\alpha:\beta]\in \mathrm{P}\mathbb{Q}^{1}$, and $R\in
\mathrm{P}\mathbb{Q}^{n-1}$. Any point in $\mathrm{P}\mathbb{Q}^{n-1}$ must
lie on such a line, and each line must intersect the surface $C$ at a single
other point, or be in $C$. Thus by cycling through $R$ (and when appropriate
$\alpha$ and $\beta$) we can parameterise all solutions to the cubic
equation. 

For those lines not in $C$ we can find the value of $[\alpha:\beta]$ by
substitution into the cubic equation. On such a substitution we obtain 
\begin{multline}
\beta^2\left\{3\alpha\sum_{i=1}^n S_i\left[R_i^2-\left(\sum_{i=1}^n R_i\right)^2\right]\right.\\ \left.+\beta\sum_{i=1}^n R_i\left[R_i^2-\left(\sum_{i=1}^n R_i\right)^2\right]\right\}=0.
\end{multline}
Since $\beta=0$ returns $S$ the other point of intersection occurs at
\begin{align}
[\alpha:\beta]=\left[\sum_{i=1}^n R_i\left\{R_i^2-\left(\sum_{i=1}^n R_i\right)^2\right\}\right.:\\\left.-3\alpha\sum_{i=1}^n S_i\left\{R_i^2-\left(\sum_{i=1}^n R_i\right)^2\right\}\right].
\end{align}
\section{Catching points on lines in the quadratic hypersurface \label{sec:catch}}
In \S\ref{sec:famNonU} we stated that the parameterisation of (\ref{genSol}) did
not catch all of the solutions:
it missed those which sit on lines $L$
through the known solution $\tilde X$,
which themselves lie in the hypersurface
$\Gamma$. Using the construction in \S\ref{sec:famNonU} with different known
solutions $\tilde X$, we can ensure that every point can be written as the
second intersection of a line, $L$, through a point $\tilde X$ for which $L
\nsubseteq \Gamma$. Every solution, $X$, for which $X^{ T}H \tilde X\ne 0$ is
covered by the known solution $\tilde X$ (as $X^{ T}H \tilde X\ne
0\Leftrightarrow R^TH\tilde X \ne 0$).  Table~\ref{setTable} lists a set of 11
pairs of 
$\tilde X,\ X$ for which $X^{ T} H \tilde X=0$. The only solution
$X$ which satisfies all the given conditions is the trivial solution. Thus,
any non-trivial solution can be obtained by (\ref{genSol}),
using an instance of $\tilde X$ from Table~\ref{setTable}. 
\renewcommand{\arraystretch}{1.5} 
\begin{table*}[t]
\begin{center}
{\small
 \begin{tabular}{|c|c|c|}
\hline
$i$ & Non-zero elements of $\tilde X$ &Condition on $X$ for $X^{T}H\tilde X=0$ \\ \hline
1 &$Q_\alpha=L_\alpha=1$ & $Q_\alpha=L_\alpha$\\ 
2 & $D_\alpha=L_\alpha=1$ & $D_\alpha=L_\alpha$\\
3 & $L_\alpha=E_\alpha=1$ & $E_\alpha=L_\alpha$\\
4 &$Q_\alpha=U_\alpha=D_\alpha=L_\alpha=E_\alpha=1$  & $Q_\alpha-2U_\alpha+D_\alpha-L_\alpha+E_\alpha=0$\\
5 &$Q_\beta=L_\beta=1$ & $Q_\beta=L_\beta$\\
6 &$D_\beta=L_\beta=1$ & $D_\beta=L_\beta$\\
7 & $L_\beta=E_\beta=1$ &$E_\beta=L_\beta$\\
8 & $Q_\beta=U_\beta=D_\beta=L_\beta=E_\beta=1$ & $Q_\beta-2U_\beta+D_\beta-L_\beta+E_\beta=0$ \\
9 &$Q_\alpha=3$, $U_\beta=1$, $L_\beta=1$ & $Q_\alpha-2 D_\beta - L_\beta=0$\\
10 &$Q_+=3$, $Q_\beta=4$ & $3Q_\beta-4Q_+=0$\\
11 &$U_+=3$, $D_\beta=-4$ & $3D_\beta+4U_+=0$\\ \hline
  \end{tabular} }
\caption{A set of known solutions $\tilde X$, which together can generate all
  non-trivial solutions to $X^THX=0$, along with a condition on $X$ such that
  $X^TH\tilde X=0$. If $X$ does {\em not}\/ satisfy this condition then it
  {\em can}\/ be generated by the corresponding $\tilde X$. } 
\label{setTable}
\end{center}
\end{table*}
\renewcommand{\arraystretch}{1} 

\section{Semi-family universal case \label{sec:semi}} 
Our choice of $GL(3,\Z)$ allows us to study the case where two families have
equal charges with relative ease, since we must set $F_\alpha$ to zero. This
has the effect of reducing the quadratic equation~(\ref{eq:quad}) to  
\begin{multline}\label{qud:newvarUniv}
2\left\{3 (Q_\beta^2-2 U_\beta^2+ D_\beta^2- L_\beta^2+ E_\beta^2)\right.\\-4Q_+(Q_\beta-2D_\beta+3L_\beta+2E_\beta)\\ \left.+4U_+(2U_\beta+D_\beta+E_\beta)\right\}=0.
\end{multline}
We can follow the same procedure as for the NFU case, except now
\begin{align}
X^T\equiv(Q_+,U_+,Q_\beta,U_\beta,D_\beta,L_\beta,E_\beta).
\end{align}
Suitable choices for $\tilde X$ may be read from $5\le i\le11$
of Table~\ref{setTable}.


\bibliography{anom}

\newcommand{\noop}[1]{}
\begin{thebibliography}{63}
\expandafter\ifx\csname natexlab\endcsname\relax\def\natexlab#1{#1}\fi
\expandafter\ifx\csname bibnamefont\endcsname\relax
  \def\bibnamefont#1{#1}\fi
\expandafter\ifx\csname bibfnamefont\endcsname\relax
  \def\bibfnamefont#1{#1}\fi
\expandafter\ifx\csname citenamefont\endcsname\relax
  \def\citenamefont#1{#1}\fi
\expandafter\ifx\csname url\endcsname\relax
  \def\url#1{\texttt{#1}}\fi
\expandafter\ifx\csname urlprefix\endcsname\relax\def\urlprefix{URL }\fi
\providecommand{\bibinfo}[2]{#2}
\providecommand{\eprint}[2][]{\url{#2}}

\bibitem[{\citenamefont{Weinberg}(1995)}]{weinberg1995quantum2}
\bibinfo{author}{\bibfnamefont{S.}~\bibnamefont{Weinberg}},
  \emph{\bibinfo{title}{The quantum theory of fields}},
  vol.~\bibinfo{volume}{2} (\bibinfo{publisher}{Cambridge university press},
  \bibinfo{year}{1995}).

\bibitem[{\citenamefont{Lohitsiri and Tong}(2019)}]{Lohitsiri:2019fuu}
\bibinfo{author}{\bibfnamefont{N.}~\bibnamefont{Lohitsiri}} \bibnamefont{and}
  \bibinfo{author}{\bibfnamefont{D.}~\bibnamefont{Tong}}
  (\bibinfo{year}{2019}), \eprint{1907.00514}.

\bibitem[{\citenamefont{Davighi et~al.}(2019)\citenamefont{Davighi, Gripaios,
  and Lohitsiri}}]{Davighi:2019rcd}
\bibinfo{author}{\bibfnamefont{J.}~\bibnamefont{Davighi}},
  \bibinfo{author}{\bibfnamefont{B.}~\bibnamefont{Gripaios}}, \bibnamefont{and}
  \bibinfo{author}{\bibfnamefont{N.}~\bibnamefont{Lohitsiri}}
  (\bibinfo{year}{2019}), \eprint{1910.11277}.

\bibitem[{\citenamefont{Davighi et~al.}()\citenamefont{Davighi, Gripaios, and
  Lohitsiri}}]{Davighi:notyet}
\bibinfo{author}{\bibfnamefont{J.}~\bibnamefont{Davighi}},
  \bibinfo{author}{\bibfnamefont{B.}~\bibnamefont{Gripaios}}, \bibnamefont{and}
  \bibinfo{author}{\bibfnamefont{N.}~\bibnamefont{Lohitsiri}} (????),
  \bibinfo{note}{to appear}.

\bibitem[{\citenamefont{Fileviez~Perez and Wise}(2010)}]{FileviezPerez:2010gw}
\bibinfo{author}{\bibfnamefont{P.}~\bibnamefont{Fileviez~Perez}}
  \bibnamefont{and} \bibinfo{author}{\bibfnamefont{M.~B.} \bibnamefont{Wise}},
  \bibinfo{journal}{Phys. Rev.} \textbf{\bibinfo{volume}{D82}},
  \bibinfo{pages}{011901} (\bibinfo{year}{2010}), \bibinfo{note}{[Erratum:
  Phys. Rev.D82,079901(2010)]}, \eprint{1002.1754}.

\bibitem[{\citenamefont{Okada and Seto}(2010)}]{Okada:2010wd}
\bibinfo{author}{\bibfnamefont{N.}~\bibnamefont{Okada}} \bibnamefont{and}
  \bibinfo{author}{\bibfnamefont{O.}~\bibnamefont{Seto}},
  \bibinfo{journal}{Phys. Rev.} \textbf{\bibinfo{volume}{D82}},
  \bibinfo{pages}{023507} (\bibinfo{year}{2010}), \eprint{1002.2525}.

\bibitem[{\citenamefont{Nakayama et~al.}(2011)\citenamefont{Nakayama,
  Takahashi, and Yanagida}}]{Nakayama:2011dj}
\bibinfo{author}{\bibfnamefont{K.}~\bibnamefont{Nakayama}},
  \bibinfo{author}{\bibfnamefont{F.}~\bibnamefont{Takahashi}},
  \bibnamefont{and} \bibinfo{author}{\bibfnamefont{T.~T.}
  \bibnamefont{Yanagida}}, \bibinfo{journal}{Phys. Lett.}
  \textbf{\bibinfo{volume}{B699}}, \bibinfo{pages}{360} (\bibinfo{year}{2011}),
  \eprint{1102.4688}.

\bibitem[{\citenamefont{Allanach et~al.}(2016)\citenamefont{Allanach, Queiroz,
  Strumia, and Sun}}]{Allanach:2015gkd}
\bibinfo{author}{\bibfnamefont{B.}~\bibnamefont{Allanach}},
  \bibinfo{author}{\bibfnamefont{F.~S.} \bibnamefont{Queiroz}},
  \bibinfo{author}{\bibfnamefont{A.}~\bibnamefont{Strumia}}, \bibnamefont{and}
  \bibinfo{author}{\bibfnamefont{S.}~\bibnamefont{Sun}},
  \bibinfo{journal}{Phys. Rev.} \textbf{\bibinfo{volume}{D93}},
  \bibinfo{pages}{055045} (\bibinfo{year}{2016}), \bibinfo{note}{[Erratum:
  Phys. Rev.D95,no.11,119902(2017)]}, \eprint{1511.07447}.

\bibitem[{\citenamefont{Okada and Okada}(2017)}]{Okada:2016tci}
\bibinfo{author}{\bibfnamefont{N.}~\bibnamefont{Okada}} \bibnamefont{and}
  \bibinfo{author}{\bibfnamefont{S.}~\bibnamefont{Okada}},
  \bibinfo{journal}{Phys. Rev.} \textbf{\bibinfo{volume}{D95}},
  \bibinfo{pages}{035025} (\bibinfo{year}{2017}), \eprint{1611.02672}.

\bibitem[{\citenamefont{Okada and Okada}(2016)}]{Okada:2016gsh}
\bibinfo{author}{\bibfnamefont{N.}~\bibnamefont{Okada}} \bibnamefont{and}
  \bibinfo{author}{\bibfnamefont{S.}~\bibnamefont{Okada}},
  \bibinfo{journal}{Phys. Rev.} \textbf{\bibinfo{volume}{D93}},
  \bibinfo{pages}{075003} (\bibinfo{year}{2016}), \eprint{1601.07526}.

\bibitem[{\citenamefont{Okada et~al.}(2018{\natexlab{a}})\citenamefont{Okada,
  Okada, and Raut}}]{Okada:2017dqs}
\bibinfo{author}{\bibfnamefont{N.}~\bibnamefont{Okada}},
  \bibinfo{author}{\bibfnamefont{S.}~\bibnamefont{Okada}}, \bibnamefont{and}
  \bibinfo{author}{\bibfnamefont{D.}~\bibnamefont{Raut}},
  \bibinfo{journal}{Phys. Lett.} \textbf{\bibinfo{volume}{B780}},
  \bibinfo{pages}{422} (\bibinfo{year}{2018}{\natexlab{a}}),
  \eprint{1712.05290}.

\bibitem[{\citenamefont{Agrawal et~al.}(2018)\citenamefont{Agrawal, Kitajima,
  Reece, Sekiguchi, and Takahashi}}]{Agrawal:2018vin}
\bibinfo{author}{\bibfnamefont{P.}~\bibnamefont{Agrawal}},
  \bibinfo{author}{\bibfnamefont{N.}~\bibnamefont{Kitajima}},
  \bibinfo{author}{\bibfnamefont{M.}~\bibnamefont{Reece}},
  \bibinfo{author}{\bibfnamefont{T.}~\bibnamefont{Sekiguchi}},
  \bibnamefont{and} \bibinfo{author}{\bibfnamefont{F.}~\bibnamefont{Takahashi}}
  (\bibinfo{year}{2018}), \eprint{1810.07188}.

\bibitem[{\citenamefont{Okada et~al.}(2018{\natexlab{b}})\citenamefont{Okada,
  Okada, and Raut}}]{Okada:2018tgy}
\bibinfo{author}{\bibfnamefont{N.}~\bibnamefont{Okada}},
  \bibinfo{author}{\bibfnamefont{S.}~\bibnamefont{Okada}}, \bibnamefont{and}
  \bibinfo{author}{\bibfnamefont{D.}~\bibnamefont{Raut}}
  (\bibinfo{year}{2018}{\natexlab{b}}), \eprint{1811.11927}.

\bibitem[{\citenamefont{Heeck and Rodejohann}(2011)}]{Heeck:2011wj}
\bibinfo{author}{\bibfnamefont{J.}~\bibnamefont{Heeck}} \bibnamefont{and}
  \bibinfo{author}{\bibfnamefont{W.}~\bibnamefont{Rodejohann}},
  \bibinfo{journal}{Phys. Rev.} \textbf{\bibinfo{volume}{D84}},
  \bibinfo{pages}{075007} (\bibinfo{year}{2011}), \eprint{1107.5238}.

\bibitem[{\citenamefont{Berenstein and Perkins}(2010)}]{Berenstein:2010ta}
\bibinfo{author}{\bibfnamefont{D.}~\bibnamefont{Berenstein}} \bibnamefont{and}
  \bibinfo{author}{\bibfnamefont{E.}~\bibnamefont{Perkins}},
  \bibinfo{journal}{Phys. Rev.} \textbf{\bibinfo{volume}{D82}},
  \bibinfo{pages}{107701} (\bibinfo{year}{2010}), \eprint{1003.4233}.

\bibitem[{\citenamefont{Chen et~al.}(2012)\citenamefont{Chen, Huang, and
  Shepherd}}]{Chen:2011sb}
\bibinfo{author}{\bibfnamefont{M.-C.} \bibnamefont{Chen}},
  \bibinfo{author}{\bibfnamefont{J.}~\bibnamefont{Huang}}, \bibnamefont{and}
  \bibinfo{author}{\bibfnamefont{W.}~\bibnamefont{Shepherd}},
  \bibinfo{journal}{JHEP} \textbf{\bibinfo{volume}{11}}, \bibinfo{pages}{059}
  (\bibinfo{year}{2012}), \eprint{1111.5018}.

\bibitem[{\citenamefont{Carone et~al.}(1996)\citenamefont{Carone, Hall, and
  Murayama}}]{Carone:1996nd}
\bibinfo{author}{\bibfnamefont{C.~D.} \bibnamefont{Carone}},
  \bibinfo{author}{\bibfnamefont{L.~J.} \bibnamefont{Hall}}, \bibnamefont{and}
  \bibinfo{author}{\bibfnamefont{H.}~\bibnamefont{Murayama}},
  \bibinfo{journal}{Phys. Rev.} \textbf{\bibinfo{volume}{D54}},
  \bibinfo{pages}{2328} (\bibinfo{year}{1996}), \eprint{hep-ph/9602364}.

\bibitem[{\citenamefont{Kaplan and Kribs}(2000)}]{Kaplan:1999iq}
\bibinfo{author}{\bibfnamefont{D.~E.} \bibnamefont{Kaplan}} \bibnamefont{and}
  \bibinfo{author}{\bibfnamefont{G.~D.} \bibnamefont{Kribs}},
  \bibinfo{journal}{Phys. Rev.} \textbf{\bibinfo{volume}{D61}},
  \bibinfo{pages}{075011} (\bibinfo{year}{2000}), \eprint{hep-ph/9906341}.

\bibitem[{\citenamefont{Froggatt and Nielsen}(1979)}]{Froggatt:1978nt}
\bibinfo{author}{\bibfnamefont{C.~D.} \bibnamefont{Froggatt}} \bibnamefont{and}
  \bibinfo{author}{\bibfnamefont{H.~B.} \bibnamefont{Nielsen}},
  \bibinfo{journal}{Nucl. Phys.} \textbf{\bibinfo{volume}{B147}},
  \bibinfo{pages}{277} (\bibinfo{year}{1979}).

\bibitem[{\citenamefont{Gauld et~al.}(2014)\citenamefont{Gauld, Goertz, and
  Haisch}}]{Gauld:2013qba}
\bibinfo{author}{\bibfnamefont{R.}~\bibnamefont{Gauld}},
  \bibinfo{author}{\bibfnamefont{F.}~\bibnamefont{Goertz}}, \bibnamefont{and}
  \bibinfo{author}{\bibfnamefont{U.}~\bibnamefont{Haisch}},
  \bibinfo{journal}{Phys. Rev.} \textbf{\bibinfo{volume}{D89}},
  \bibinfo{pages}{015005} (\bibinfo{year}{2014}), \eprint{1308.1959}.

\bibitem[{\citenamefont{Buras et~al.}(2014{\natexlab{a}})\citenamefont{Buras,
  De~Fazio, and Girrbach}}]{Buras:2013dea}
\bibinfo{author}{\bibfnamefont{A.~J.} \bibnamefont{Buras}},
  \bibinfo{author}{\bibfnamefont{F.}~\bibnamefont{De~Fazio}}, \bibnamefont{and}
  \bibinfo{author}{\bibfnamefont{J.}~\bibnamefont{Girrbach}},
  \bibinfo{journal}{JHEP} \textbf{\bibinfo{volume}{02}}, \bibinfo{pages}{112}
  (\bibinfo{year}{2014}{\natexlab{a}}), \eprint{1311.6729}.

\bibitem[{\citenamefont{Buras and Girrbach}(2013)}]{Buras:2013qja}
\bibinfo{author}{\bibfnamefont{A.~J.} \bibnamefont{Buras}} \bibnamefont{and}
  \bibinfo{author}{\bibfnamefont{J.}~\bibnamefont{Girrbach}},
  \bibinfo{journal}{JHEP} \textbf{\bibinfo{volume}{12}}, \bibinfo{pages}{009}
  (\bibinfo{year}{2013}), \eprint{1309.2466}.

\bibitem[{\citenamefont{Altmannshofer et~al.}(2014)\citenamefont{Altmannshofer,
  Gori, Pospelov, and Yavin}}]{Altmannshofer:2014cfa}
\bibinfo{author}{\bibfnamefont{W.}~\bibnamefont{Altmannshofer}},
  \bibinfo{author}{\bibfnamefont{S.}~\bibnamefont{Gori}},
  \bibinfo{author}{\bibfnamefont{M.}~\bibnamefont{Pospelov}}, \bibnamefont{and}
  \bibinfo{author}{\bibfnamefont{I.}~\bibnamefont{Yavin}},
  \bibinfo{journal}{Phys. Rev.} \textbf{\bibinfo{volume}{D89}},
  \bibinfo{pages}{095033} (\bibinfo{year}{2014}), \eprint{1403.1269}.

\bibitem[{\citenamefont{Buras et~al.}(2014{\natexlab{b}})\citenamefont{Buras,
  De~Fazio, and Girrbach-Noe}}]{Buras:2014yna}
\bibinfo{author}{\bibfnamefont{A.~J.} \bibnamefont{Buras}},
  \bibinfo{author}{\bibfnamefont{F.}~\bibnamefont{De~Fazio}}, \bibnamefont{and}
  \bibinfo{author}{\bibfnamefont{J.}~\bibnamefont{Girrbach-Noe}},
  \bibinfo{journal}{JHEP} \textbf{\bibinfo{volume}{08}}, \bibinfo{pages}{039}
  (\bibinfo{year}{2014}{\natexlab{b}}), \eprint{1405.3850}.

\bibitem[{\citenamefont{Crivellin
  et~al.}(2015{\natexlab{a}})\citenamefont{Crivellin, D'Ambrosio, and
  Heeck}}]{Crivellin:2015mga}
\bibinfo{author}{\bibfnamefont{A.}~\bibnamefont{Crivellin}},
  \bibinfo{author}{\bibfnamefont{G.}~\bibnamefont{D'Ambrosio}},
  \bibnamefont{and} \bibinfo{author}{\bibfnamefont{J.}~\bibnamefont{Heeck}},
  \bibinfo{journal}{Phys. Rev. Lett.} \textbf{\bibinfo{volume}{114}},
  \bibinfo{pages}{151801} (\bibinfo{year}{2015}{\natexlab{a}}),
  \eprint{1501.00993}.

\bibitem[{\citenamefont{Crivellin
  et~al.}(2015{\natexlab{b}})\citenamefont{Crivellin, D'Ambrosio, and
  Heeck}}]{Crivellin:2015lwa}
\bibinfo{author}{\bibfnamefont{A.}~\bibnamefont{Crivellin}},
  \bibinfo{author}{\bibfnamefont{G.}~\bibnamefont{D'Ambrosio}},
  \bibnamefont{and} \bibinfo{author}{\bibfnamefont{J.}~\bibnamefont{Heeck}},
  \bibinfo{journal}{Phys. Rev.} \textbf{\bibinfo{volume}{D91}},
  \bibinfo{pages}{075006} (\bibinfo{year}{2015}{\natexlab{b}}),
  \eprint{1503.03477}.

\bibitem[{\citenamefont{Aristizabal~Sierra
  et~al.}(2015)\citenamefont{Aristizabal~Sierra, Staub, and
  Vicente}}]{Sierra:2015fma}
\bibinfo{author}{\bibfnamefont{D.}~\bibnamefont{Aristizabal~Sierra}},
  \bibinfo{author}{\bibfnamefont{F.}~\bibnamefont{Staub}}, \bibnamefont{and}
  \bibinfo{author}{\bibfnamefont{A.}~\bibnamefont{Vicente}},
  \bibinfo{journal}{Phys. Rev.} \textbf{\bibinfo{volume}{D92}},
  \bibinfo{pages}{015001} (\bibinfo{year}{2015}), \eprint{1503.06077}.

\bibitem[{\citenamefont{Crivellin
  et~al.}(2015{\natexlab{c}})\citenamefont{Crivellin, Hofer, Matias, Nierste,
  Pokorski, and Rosiek}}]{Crivellin:2015era}
\bibinfo{author}{\bibfnamefont{A.}~\bibnamefont{Crivellin}},
  \bibinfo{author}{\bibfnamefont{L.}~\bibnamefont{Hofer}},
  \bibinfo{author}{\bibfnamefont{J.}~\bibnamefont{Matias}},
  \bibinfo{author}{\bibfnamefont{U.}~\bibnamefont{Nierste}},
  \bibinfo{author}{\bibfnamefont{S.}~\bibnamefont{Pokorski}}, \bibnamefont{and}
  \bibinfo{author}{\bibfnamefont{J.}~\bibnamefont{Rosiek}},
  \bibinfo{journal}{Phys. Rev.} \textbf{\bibinfo{volume}{D92}},
  \bibinfo{pages}{054013} (\bibinfo{year}{2015}{\natexlab{c}}),
  \eprint{1504.07928}.

\bibitem[{\citenamefont{Celis et~al.}(2015)\citenamefont{Celis, Fuentes-Martin,
  Jung, and Serodio}}]{Celis:2015ara}
\bibinfo{author}{\bibfnamefont{A.}~\bibnamefont{Celis}},
  \bibinfo{author}{\bibfnamefont{J.}~\bibnamefont{Fuentes-Martin}},
  \bibinfo{author}{\bibfnamefont{M.}~\bibnamefont{Jung}}, \bibnamefont{and}
  \bibinfo{author}{\bibfnamefont{H.}~\bibnamefont{Serodio}},
  \bibinfo{journal}{Phys. Rev.} \textbf{\bibinfo{volume}{D92}},
  \bibinfo{pages}{015007} (\bibinfo{year}{2015}), \eprint{1505.03079}.

\bibitem[{\citenamefont{Greljo et~al.}(2015)\citenamefont{Greljo, Isidori, and
  Marzocca}}]{Greljo:2015mma}
\bibinfo{author}{\bibfnamefont{A.}~\bibnamefont{Greljo}},
  \bibinfo{author}{\bibfnamefont{G.}~\bibnamefont{Isidori}}, \bibnamefont{and}
  \bibinfo{author}{\bibfnamefont{D.}~\bibnamefont{Marzocca}},
  \bibinfo{journal}{JHEP} \textbf{\bibinfo{volume}{07}}, \bibinfo{pages}{142}
  (\bibinfo{year}{2015}), \eprint{1506.01705}.

\bibitem[{\citenamefont{Altmannshofer and Yavin}(2015)}]{Altmannshofer:2015mqa}
\bibinfo{author}{\bibfnamefont{W.}~\bibnamefont{Altmannshofer}}
  \bibnamefont{and} \bibinfo{author}{\bibfnamefont{I.}~\bibnamefont{Yavin}},
  \bibinfo{journal}{Phys. Rev.} \textbf{\bibinfo{volume}{D92}},
  \bibinfo{pages}{075022} (\bibinfo{year}{2015}), \eprint{1508.07009}.

\bibitem[{\citenamefont{Falkowski et~al.}(2015)\citenamefont{Falkowski,
  Nardecchia, and Ziegler}}]{Falkowski:2015zwa}
\bibinfo{author}{\bibfnamefont{A.}~\bibnamefont{Falkowski}},
  \bibinfo{author}{\bibfnamefont{M.}~\bibnamefont{Nardecchia}},
  \bibnamefont{and} \bibinfo{author}{\bibfnamefont{R.}~\bibnamefont{Ziegler}},
  \bibinfo{journal}{JHEP} \textbf{\bibinfo{volume}{11}}, \bibinfo{pages}{173}
  (\bibinfo{year}{2015}), \eprint{1509.01249}.

\bibitem[{\citenamefont{Chiang et~al.}(2016)\citenamefont{Chiang, He, and
  Valencia}}]{Chiang:2016qov}
\bibinfo{author}{\bibfnamefont{C.-W.} \bibnamefont{Chiang}},
  \bibinfo{author}{\bibfnamefont{X.-G.} \bibnamefont{He}}, \bibnamefont{and}
  \bibinfo{author}{\bibfnamefont{G.}~\bibnamefont{Valencia}},
  \bibinfo{journal}{Phys. Rev.} \textbf{\bibinfo{volume}{D93}},
  \bibinfo{pages}{074003} (\bibinfo{year}{2016}), \eprint{1601.07328}.

\bibitem[{\citenamefont{Bečirević et~al.}(2016)\citenamefont{Bečirević,
  Sumensari, and Zukanovich~Funchal}}]{Becirevic:2016zri}
\bibinfo{author}{\bibfnamefont{D.}~\bibnamefont{Bečirević}},
  \bibinfo{author}{\bibfnamefont{O.}~\bibnamefont{Sumensari}},
  \bibnamefont{and}
  \bibinfo{author}{\bibfnamefont{R.}~\bibnamefont{Zukanovich~Funchal}},
  \bibinfo{journal}{Eur. Phys. J.} \textbf{\bibinfo{volume}{C76}},
  \bibinfo{pages}{134} (\bibinfo{year}{2016}), \eprint{1602.00881}.

\bibitem[{\citenamefont{Boucenna
  et~al.}(2016{\natexlab{a}})\citenamefont{Boucenna, Celis, Fuentes-Martin,
  Vicente, and Virto}}]{Boucenna:2016wpr}
\bibinfo{author}{\bibfnamefont{S.~M.} \bibnamefont{Boucenna}},
  \bibinfo{author}{\bibfnamefont{A.}~\bibnamefont{Celis}},
  \bibinfo{author}{\bibfnamefont{J.}~\bibnamefont{Fuentes-Martin}},
  \bibinfo{author}{\bibfnamefont{A.}~\bibnamefont{Vicente}}, \bibnamefont{and}
  \bibinfo{author}{\bibfnamefont{J.}~\bibnamefont{Virto}},
  \bibinfo{journal}{Phys. Lett.} \textbf{\bibinfo{volume}{B760}},
  \bibinfo{pages}{214} (\bibinfo{year}{2016}{\natexlab{a}}),
  \eprint{1604.03088}.

\bibitem[{\citenamefont{Boucenna
  et~al.}(2016{\natexlab{b}})\citenamefont{Boucenna, Celis, Fuentes-Martin,
  Vicente, and Virto}}]{Boucenna:2016qad}
\bibinfo{author}{\bibfnamefont{S.~M.} \bibnamefont{Boucenna}},
  \bibinfo{author}{\bibfnamefont{A.}~\bibnamefont{Celis}},
  \bibinfo{author}{\bibfnamefont{J.}~\bibnamefont{Fuentes-Martin}},
  \bibinfo{author}{\bibfnamefont{A.}~\bibnamefont{Vicente}}, \bibnamefont{and}
  \bibinfo{author}{\bibfnamefont{J.}~\bibnamefont{Virto}},
  \bibinfo{journal}{JHEP} \textbf{\bibinfo{volume}{12}}, \bibinfo{pages}{059}
  (\bibinfo{year}{2016}{\natexlab{b}}), \eprint{1608.01349}.

\bibitem[{\citenamefont{Ko et~al.}(2017)\citenamefont{Ko, Omura, Shigekami, and
  Yu}}]{Ko:2017lzd}
\bibinfo{author}{\bibfnamefont{P.}~\bibnamefont{Ko}},
  \bibinfo{author}{\bibfnamefont{Y.}~\bibnamefont{Omura}},
  \bibinfo{author}{\bibfnamefont{Y.}~\bibnamefont{Shigekami}},
  \bibnamefont{and} \bibinfo{author}{\bibfnamefont{C.}~\bibnamefont{Yu}},
  \bibinfo{journal}{Phys. Rev.} \textbf{\bibinfo{volume}{D95}},
  \bibinfo{pages}{115040} (\bibinfo{year}{2017}), \eprint{1702.08666}.

\bibitem[{\citenamefont{Alonso et~al.}(2017{\natexlab{a}})\citenamefont{Alonso,
  Cox, Han, and Yanagida}}]{Alonso:2017bff}
\bibinfo{author}{\bibfnamefont{R.}~\bibnamefont{Alonso}},
  \bibinfo{author}{\bibfnamefont{P.}~\bibnamefont{Cox}},
  \bibinfo{author}{\bibfnamefont{C.}~\bibnamefont{Han}}, \bibnamefont{and}
  \bibinfo{author}{\bibfnamefont{T.~T.} \bibnamefont{Yanagida}},
  \bibinfo{journal}{Phys. Rev.} \textbf{\bibinfo{volume}{D96}},
  \bibinfo{pages}{071701} (\bibinfo{year}{2017}{\natexlab{a}}),
  \eprint{1704.08158}.

\bibitem[{\citenamefont{Alonso et~al.}(2017{\natexlab{b}})\citenamefont{Alonso,
  Cox, Han, and Yanagida}}]{Alonso:2017uky}
\bibinfo{author}{\bibfnamefont{R.}~\bibnamefont{Alonso}},
  \bibinfo{author}{\bibfnamefont{P.}~\bibnamefont{Cox}},
  \bibinfo{author}{\bibfnamefont{C.}~\bibnamefont{Han}}, \bibnamefont{and}
  \bibinfo{author}{\bibfnamefont{T.~T.} \bibnamefont{Yanagida}},
  \bibinfo{journal}{Phys. Lett.} \textbf{\bibinfo{volume}{B774}},
  \bibinfo{pages}{643} (\bibinfo{year}{2017}{\natexlab{b}}),
  \eprint{1705.03858}.

\bibitem[{\citenamefont{Tang and Wu}(2018)}]{1674-1137-42-3-033104}
\bibinfo{author}{\bibfnamefont{Y.}~\bibnamefont{Tang}} \bibnamefont{and}
  \bibinfo{author}{\bibfnamefont{Y.-L.} \bibnamefont{Wu}},
  \bibinfo{journal}{Chinese Physics C} \textbf{\bibinfo{volume}{42}},
  \bibinfo{pages}{033104} (\bibinfo{year}{2018}),
  \urlprefix\url{http://stacks.iop.org/1674-1137/42/i=3/a=033104}.

\bibitem[{\citenamefont{Chen and Nomura}(2018)}]{CHEN2018420}
\bibinfo{author}{\bibfnamefont{C.-H.} \bibnamefont{Chen}} \bibnamefont{and}
  \bibinfo{author}{\bibfnamefont{T.}~\bibnamefont{Nomura}},
  \bibinfo{journal}{Physics Letters B} \textbf{\bibinfo{volume}{777}},
  \bibinfo{pages}{420 } (\bibinfo{year}{2018}), ISSN \bibinfo{issn}{0370-2693}.

\bibitem[{\citenamefont{Faisel and Tandean}(2018)}]{Faisel:2017glo}
\bibinfo{author}{\bibfnamefont{G.}~\bibnamefont{Faisel}} \bibnamefont{and}
  \bibinfo{author}{\bibfnamefont{J.}~\bibnamefont{Tandean}},
  \bibinfo{journal}{JHEP} \textbf{\bibinfo{volume}{02}}, \bibinfo{pages}{074}
  (\bibinfo{year}{2018}), \eprint{1710.11102}.

\bibitem[{\citenamefont{Fuyuto et~al.}(2018)\citenamefont{Fuyuto, Li, and
  Yu}}]{PhysRevD.97.115003}
\bibinfo{author}{\bibfnamefont{K.}~\bibnamefont{Fuyuto}},
  \bibinfo{author}{\bibfnamefont{H.-L.} \bibnamefont{Li}}, \bibnamefont{and}
  \bibinfo{author}{\bibfnamefont{J.-H.} \bibnamefont{Yu}},
  \bibinfo{journal}{Phys. Rev. D} \textbf{\bibinfo{volume}{97}},
  \bibinfo{pages}{115003} (\bibinfo{year}{2018}),
  \urlprefix\url{https://link.aps.org/doi/10.1103/PhysRevD.97.115003}.

\bibitem[{\citenamefont{Bian et~al.}(2018)\citenamefont{Bian, Lee, and
  Park}}]{Bian:2017xzg}
\bibinfo{author}{\bibfnamefont{L.}~\bibnamefont{Bian}},
  \bibinfo{author}{\bibfnamefont{H.~M.} \bibnamefont{Lee}}, \bibnamefont{and}
  \bibinfo{author}{\bibfnamefont{C.~B.} \bibnamefont{Park}},
  \bibinfo{journal}{Eur. Phys. J.} \textbf{\bibinfo{volume}{C78}},
  \bibinfo{pages}{306} (\bibinfo{year}{2018}), \eprint{1711.08930}.

\bibitem[{\citenamefont{Abdullah et~al.}(2018)\citenamefont{Abdullah,
  Dalchenko, Dutta, Eusebi, Huang, Kamon, Rathjens, and
  Thompson}}]{PhysRevD.97.075035}
\bibinfo{author}{\bibfnamefont{M.}~\bibnamefont{Abdullah}},
  \bibinfo{author}{\bibfnamefont{M.}~\bibnamefont{Dalchenko}},
  \bibinfo{author}{\bibfnamefont{B.}~\bibnamefont{Dutta}},
  \bibinfo{author}{\bibfnamefont{R.}~\bibnamefont{Eusebi}},
  \bibinfo{author}{\bibfnamefont{P.}~\bibnamefont{Huang}},
  \bibinfo{author}{\bibfnamefont{T.}~\bibnamefont{Kamon}},
  \bibinfo{author}{\bibfnamefont{D.}~\bibnamefont{Rathjens}}, \bibnamefont{and}
  \bibinfo{author}{\bibfnamefont{A.}~\bibnamefont{Thompson}},
  \bibinfo{journal}{Phys. Rev. D} \textbf{\bibinfo{volume}{97}},
  \bibinfo{pages}{075035} (\bibinfo{year}{2018}),
  \urlprefix\url{https://link.aps.org/doi/10.1103/PhysRevD.97.075035}.

\bibitem[{\citenamefont{Bhatia et~al.}(2017)\citenamefont{Bhatia, Chakraborty,
  and Dighe}}]{Bhatia:2017tgo}
\bibinfo{author}{\bibfnamefont{D.}~\bibnamefont{Bhatia}},
  \bibinfo{author}{\bibfnamefont{S.}~\bibnamefont{Chakraborty}},
  \bibnamefont{and} \bibinfo{author}{\bibfnamefont{A.}~\bibnamefont{Dighe}},
  \bibinfo{journal}{JHEP} \textbf{\bibinfo{volume}{03}}, \bibinfo{pages}{117}
  (\bibinfo{year}{2017}), \eprint{1701.05825}.

\bibitem[{\citenamefont{Allanach
  et~al.}(2018{\natexlab{a}})\citenamefont{Allanach, Gripaios, and
  You}}]{Allanach:2017bta}
\bibinfo{author}{\bibfnamefont{B.~C.} \bibnamefont{Allanach}},
  \bibinfo{author}{\bibfnamefont{B.}~\bibnamefont{Gripaios}}, \bibnamefont{and}
  \bibinfo{author}{\bibfnamefont{T.}~\bibnamefont{You}},
  \bibinfo{journal}{JHEP} \textbf{\bibinfo{volume}{03}}, \bibinfo{pages}{021}
  (\bibinfo{year}{2018}{\natexlab{a}}), \eprint{1710.06363}.

\bibitem[{\citenamefont{Allanach
  et~al.}(2018{\natexlab{b}})\citenamefont{Allanach, Corbett, Dolan, and
  You}}]{Allanach:2018odd}
\bibinfo{author}{\bibfnamefont{B.~C.} \bibnamefont{Allanach}},
  \bibinfo{author}{\bibfnamefont{T.}~\bibnamefont{Corbett}},
  \bibinfo{author}{\bibfnamefont{M.~J.} \bibnamefont{Dolan}}, \bibnamefont{and}
  \bibinfo{author}{\bibfnamefont{T.}~\bibnamefont{You}}
  (\bibinfo{year}{2018}{\natexlab{b}}), \eprint{1810.02166}.

\bibitem[{\citenamefont{Allanach and Davighi}(2018)}]{Allanach:2018lvl}
\bibinfo{author}{\bibfnamefont{B.}~\bibnamefont{Allanach}} \bibnamefont{and}
  \bibinfo{author}{\bibfnamefont{J.}~\bibnamefont{Davighi}}
  (\bibinfo{year}{2018}), \eprint{1809.01158}.

\bibitem[{\citenamefont{Duan et~al.}(2018)\citenamefont{Duan, Fan, Frank, Han,
  and Yang}}]{Duan:2018akc}
\bibinfo{author}{\bibfnamefont{G.~H.} \bibnamefont{Duan}},
  \bibinfo{author}{\bibfnamefont{X.}~\bibnamefont{Fan}},
  \bibinfo{author}{\bibfnamefont{M.}~\bibnamefont{Frank}},
  \bibinfo{author}{\bibfnamefont{C.}~\bibnamefont{Han}}, \bibnamefont{and}
  \bibinfo{author}{\bibfnamefont{J.~M.} \bibnamefont{Yang}}
  (\bibinfo{year}{2018}), \eprint{1808.04116}.

\bibitem[{\citenamefont{Geng and Okada}(2018)}]{Geng:2018xzd}
\bibinfo{author}{\bibfnamefont{C.-Q.} \bibnamefont{Geng}} \bibnamefont{and}
  \bibinfo{author}{\bibfnamefont{H.}~\bibnamefont{Okada}}
  (\bibinfo{year}{2018}), \eprint{1812.07918}.

\bibitem[{\citenamefont{Kawamura et~al.}(2019)\citenamefont{Kawamura, Raby, and
  Trautner}}]{Kawamura:2019rth}
\bibinfo{author}{\bibfnamefont{J.}~\bibnamefont{Kawamura}},
  \bibinfo{author}{\bibfnamefont{S.}~\bibnamefont{Raby}}, \bibnamefont{and}
  \bibinfo{author}{\bibfnamefont{A.}~\bibnamefont{Trautner}}
  (\bibinfo{year}{2019}), \eprint{1906.11297}.

\bibitem[{\citenamefont{Allanach
  et~al.}(2019{\natexlab{a}})\citenamefont{Allanach, Butterworth, and
  Corbett}}]{Allanach:2019mfl}
\bibinfo{author}{\bibfnamefont{B.~C.} \bibnamefont{Allanach}},
  \bibinfo{author}{\bibfnamefont{J.~M.} \bibnamefont{Butterworth}},
  \bibnamefont{and} \bibinfo{author}{\bibfnamefont{T.}~\bibnamefont{Corbett}}
  (\bibinfo{year}{2019}{\natexlab{a}}), \eprint{1904.10954}.

\bibitem[{\citenamefont{Dwivedi et~al.}(2019)\citenamefont{Dwivedi, Falkowski,
  Ghosh, and Ghosh}}]{Dwivedi:2019uqd}
\bibinfo{author}{\bibfnamefont{S.}~\bibnamefont{Dwivedi}},
  \bibinfo{author}{\bibfnamefont{A.}~\bibnamefont{Falkowski}},
  \bibinfo{author}{\bibfnamefont{D.~K.} \bibnamefont{Ghosh}}, \bibnamefont{and}
  \bibinfo{author}{\bibfnamefont{N.}~\bibnamefont{Ghosh}}
  (\bibinfo{year}{2019}), \eprint{1908.03031}.

\bibitem[{\citenamefont{Allanach and Davighi}(2019)}]{Allanach:2019iiy}
\bibinfo{author}{\bibfnamefont{B.~C.} \bibnamefont{Allanach}} \bibnamefont{and}
  \bibinfo{author}{\bibfnamefont{J.}~\bibnamefont{Davighi}}
  (\bibinfo{year}{2019}), \eprint{1905.10327}.

\bibitem[{\citenamefont{Altmannshofer et~al.}(2019)\citenamefont{Altmannshofer,
  Davighi, and Nardecchia}}]{Altmannshofer:2019xda}
\bibinfo{author}{\bibfnamefont{W.}~\bibnamefont{Altmannshofer}},
  \bibinfo{author}{\bibfnamefont{J.}~\bibnamefont{Davighi}}, \bibnamefont{and}
  \bibinfo{author}{\bibfnamefont{M.}~\bibnamefont{Nardecchia}}
  (\bibinfo{year}{2019}), \eprint{1909.02021}.

\bibitem[{\citenamefont{Calibbi et~al.}(2019)\citenamefont{Calibbi, Crivellin,
  Kirk, Manzari, and Vernazza}}]{Calibbi:2019lvs}
\bibinfo{author}{\bibfnamefont{L.}~\bibnamefont{Calibbi}},
  \bibinfo{author}{\bibfnamefont{A.}~\bibnamefont{Crivellin}},
  \bibinfo{author}{\bibfnamefont{F.}~\bibnamefont{Kirk}},
  \bibinfo{author}{\bibfnamefont{C.~A.} \bibnamefont{Manzari}},
  \bibnamefont{and} \bibinfo{author}{\bibfnamefont{L.}~\bibnamefont{Vernazza}}
  (\bibinfo{year}{2019}), \eprint{1910.00014}.

\bibitem[{\citenamefont{Aaij et~al.}(2014)}]{Aaij:2014ora}
\bibinfo{author}{\bibfnamefont{R.}~\bibnamefont{Aaij}} \bibnamefont{et~al.}
  (\bibinfo{collaboration}{LHCb}), \bibinfo{journal}{Phys. Rev. Lett.}
  \textbf{\bibinfo{volume}{113}}, \bibinfo{pages}{151601}
  (\bibinfo{year}{2014}), \eprint{1406.6482}.

\bibitem[{\citenamefont{Aaij et~al.}(2017)}]{Aaij:2017vbb}
\bibinfo{author}{\bibfnamefont{R.}~\bibnamefont{Aaij}} \bibnamefont{et~al.}
  (\bibinfo{collaboration}{LHCb}), \bibinfo{journal}{JHEP}
  \textbf{\bibinfo{volume}{08}}, \bibinfo{pages}{055} (\bibinfo{year}{2017}),
  \eprint{1705.05802}.

\bibitem[{\citenamefont{Hiller and Kruger}(2004)}]{Hiller:2003js}
\bibinfo{author}{\bibfnamefont{G.}~\bibnamefont{Hiller}} \bibnamefont{and}
  \bibinfo{author}{\bibfnamefont{F.}~\bibnamefont{Kruger}},
  \bibinfo{journal}{Phys. Rev.} \textbf{\bibinfo{volume}{D69}},
  \bibinfo{pages}{074020} (\bibinfo{year}{2004}), \eprint{hep-ph/0310219}.

\bibitem[{\citenamefont{Allanach
  et~al.}(2019{\natexlab{b}})\citenamefont{Allanach, Davighi, and
  Melville}}]{Allanach:2018vjg}
\bibinfo{author}{\bibfnamefont{B.~C.} \bibnamefont{Allanach}},
  \bibinfo{author}{\bibfnamefont{J.}~\bibnamefont{Davighi}}, \bibnamefont{and}
  \bibinfo{author}{\bibfnamefont{S.}~\bibnamefont{Melville}},
  \bibinfo{journal}{JHEP} \textbf{\bibinfo{volume}{02}}, \bibinfo{pages}{082}
  (\bibinfo{year}{2019}{\natexlab{b}}), \eprint{1812.04602}.

\bibitem[{\citenamefont{Costa et~al.}(2019)\citenamefont{Costa, Dobrescu, and
  Fox}}]{Costa_Dobrescu_Fox_2019}
\bibinfo{author}{\bibfnamefont{D.~B.} \bibnamefont{Costa}},
  \bibinfo{author}{\bibfnamefont{B.~A.} \bibnamefont{Dobrescu}},
  \bibnamefont{and} \bibinfo{author}{\bibfnamefont{P.~J.} \bibnamefont{Fox}},
  \bibinfo{journal}{Physical Review Letters} \textbf{\bibinfo{volume}{123}},
  \bibinfo{pages}{151601} (\bibinfo{year}{2019}), ISSN
  \bibinfo{issn}{0031-9007, 1079-7114}.

\bibitem[{\citenamefont{Allanach
  et~al.}(2019{\natexlab{c}})\citenamefont{Allanach, Gripaios, and
  Tooby-Smith}}]{Allanach:2019gwp}
\bibinfo{author}{\bibfnamefont{B.~C.} \bibnamefont{Allanach}},
  \bibinfo{author}{\bibfnamefont{B.}~\bibnamefont{Gripaios}}, \bibnamefont{and}
  \bibinfo{author}{\bibfnamefont{J.}~\bibnamefont{Tooby-Smith}}
  (\bibinfo{year}{2019}{\natexlab{c}}), \eprint{1912.04804}.

\end{thebibliography}

\end{document}